\def\ll_lsun{log$({L/\rm L_{\odot}})$~}  
  
\def\masa_msun{$M/ \rm M_{\odot}$~}  
  
\def\m_mstar{$M/M_{*}$~}

\def\pg{\mbox{PG~0122+200}}  
\def\pp{\mbox{PG~1159$-$035}}  
\def\pr{\mbox{PG~2131+066}}  
\def\pt{\mbox{PG~1707+427}}  
\def\rxj{\mbox{RX~J2117.1+3412}}  
\def\v4334{\mbox{V4334 Sgr}}  
\def\ngc{\mbox{NGC 1501}}  
\def\vv{\mbox{VV 47}}  
\def\J0{\mbox{SDSS J0349$-$0059}}  
\def\rrr{\mbox{SDSS J0754+0852}}

\documentclass[structabstract]{aa}
\usepackage{graphicx}  
\usepackage{txfonts}
\usepackage{subfigure}
\usepackage{natbib}
\bibpunct{(}{)}{;}{a}{}{,} % to follow the A&A style
\begin{document}  
  
\title{Asteroseismology of the GW Virginis stars \J0\ and \vv}  

\author{Leila M. Calcaferro\inst{1,2}, 
        Alejandro H. C\'orsico\inst{1,2}, \and 
        Leandro G. Althaus\inst{1,2}}  
\offprints{L. M. Calcaferro} 

\institute{$^1$ Grupo  de Evoluci\'on  Estelar y  Pulsaciones,  Facultad de 
           Ciencias Astron\'omicas  y Geof\'{\i}sicas, Universidad  Nacional de
           La Plata, Paseo del Bosque s/n, (1900) La Plata, Argentina\\  
           $^{2}$ Instituto de Astrof\'{\i}sica La Plata, CONICET-UNLP, Paseo 
           del Bosque s/n, (1900) La Plata, Argentina\\ 
           \email{lcalcaferro,acorsico,althaus@fcaglp.unlp.edu.ar}}
\date{Received ; accepted}  

\abstract{GW Virginis stars are a  well-studied class of  
non-radial $g$-mode pulsators.  
\J0\  and  \vv\ are two PG 1159 stars  
members of this class of variable stars. 
\J0\ is an interesting GW Vir star that shows a complete pulsation 
spectrum, that includes rotational splitting of some of its frequencies.  
\vv\ is  a pulsating PG 1159 star 
surrounded by a planetary  nebula. This star is particularly  
interesting  because  it  exhibits a  rich  and  complex pulsation  
spectrum.}{We present an asteroseismological study
of \J0\ and \vv\ aimed mainly at deriving their total mass on  
the basis  of state-of-the-art PG 1159  
evolutionary models.}{We  compute adiabatic  nonradial $g$-mode
pulsation  periods  for PG 1159 evolutionary models with stellar masses
ranging from $0.515$ to $0.741\ M_{\sun}$, that take into  
account the complete  evolution of the progenitor  stars.  
We first estimate  a mean period spacing for
both  \J0\ and \vv, and then we
constrain   the stellar  mass of  these stars  by comparing  the
observed period  spacing  with  the asymptotic  period spacing and
with the average of the computed period spacings.  We also employ
the  individual   observed   periods  to   search  for   a
representative seismological model for each star. Finally, we
estimate the rotation period of \J0.}{We found a spectroscopic
mass  of $M_{\star}\sim 0.543 M_{\sun}$ for \J0\ and 
$M_{\star}\sim 0.529\ M_{\sun}$ for \vv. By comparing the observed
period  spacing  with  the asymptotic  period spacing we obtain
$M_{\star}\sim 0.569\ M_{\sun}$ for \J0\ and $M_{\star}\sim 
0.523\ M_{\sun}$ for \vv.  If we compare the  observed period 
spacing with the average  of the computed period spacings we 
found $M_{\star}\sim 0.535\ M_{\sun}$ for \J0 and 
$M_{\star}\sim  0.528 M_{\sun}$ for \vv. Searching for the best period 
fit we found, in the case of 
\J0, an asteroseismological model with $M_{\star}= 0.542\ M_{\sun}$ 
and $T_{\rm eff}= 91\, 255\ $K. 
For \vv, we could not find a unique and unambiguous 
asteroseismological model. Finally,
 for \J0, we determined  the rotation period of \J0 as being 
$P_{\rm rot}= 1/\Omega \sim 0.407$ days.}{The results
presented in this work constitute a further step in the study
of GW Vir stars through asteroseismology in the frame  
of fully evolutionary models of PG 1159 stars. In particular, 
once again it is shown the potential of asteroseismology 
to derive stellar masses of PG 1159 stars with an unprecedented 
precision.} 
\titlerunning{Asteroseismology of \J0 and \vv}  
\maketitle  
\authorrunning{Calcaferro et al.}  

%----------------------------------------------------------------  
   
\section{Introduction}  
\label{intro}  

GW  Virginis stars include  pulsating stars characterized by 
several spectral types \citep[see][]{2007ApJS..171..219Q}. 
Among them, the PG 1159 stars,  are very hot
H-deficient  post-Asymptotic  Giant  Branch  (AGB)  stars  with
surface layers  rich in  He, C and
O \citep{2006PASP..118..183W,2014A&A...564A..53W,2015A&A...584A..19W}.
Pulsating PG 1159 stars  exhibit multiperiodic luminosity variations  
with periods ranging
from 5 to  100 minutes, attributable to  non-radial pulsation $g$
modes. Some GW Vir are still embedded  in a nebula \citep[see
the reviews
by][]{2008ARA&A..46..157W,2008PASP..120.1043F,2010A&ARv..18..471A}\footnote{For historical
reasons, some authors designate Planetary Nebula Nuclei
Variable (PNNV) to GW Vir stars still embedded in a nebula, and DOV to
GW Vir stars without planetary nebula \citep{2008ARA&A..46..157W}, a
misnomer because
no white dwarf of spectral type DO have ever been found to pulsate.}.  
PG 1159 stars are thought
to be  the evolutionary link between Wolf-Rayet type central stars of
planetary  nebulae and most of the H-deficient white
dwarfs \citep{1985ApJS...58..379W,1986PASP...98..821S,2005A&A...435..631A}.
It  is generally  accepted that these stars  have their  origin in a
born-again episode induced  by a post-AGB He thermal  pulse
---see \citet{1983ApJ...264..605I,1999A&A...349L...5H,2003ApJ...583..913L,
2005A&A...435..631A,2006A&A...454..845M} for references. 

Notably, considerable observational effort has been invested to study
GW Vir  stars. Particularly  noteworthy  are  the works
of \citet{2002A&A...381..122V} on \rxj, \citet{2007A&A...467..237F}
on \pg,  and \citet{2008A&A...477..627C}
and \citet{2008A&A...489.1225C} on \pp.  On  the  theoretical front,
important  progress in the  numerical  modeling of  PG 1159
stars \citep{2005A&A...435..631A,
2006A&A...454..845M,2007A&A...470..675M,2007MNRAS.380..763M} has paved
the way for unprecedented asteroseismological inferences for the
mentioned stars 
\citep{2007A&A...461.1095C,2007A&A...475..619C,2008A&A...478..869C},
and also for \pr, \pt, \ngc, and \rrr\ 
\citep{2009A&A...499..257C,2014MNRAS.442.2278K}.  The
detailed PG 1159 stellar models of \cite{2006A&A...454..845M}  were
derived  from the  complete evolutionary  history of progenitor stars
with different stellar masses and an elaborate treatment of the mixing
and  extra-mixing  processes  during   the core  He  burning  and
born-again  phases.  It  is  worth mentioning  that  these models  are
characterized by  thick helium-rich  outer envelopes. The
robustness of the H-deficient post-AGB tracks
of \cite{2006A&A...454..845M} regarding  previous evolution of their
progenitor stars and the constitutive  physics of the remnants have
been assesed by \citet{2007A&A...470..675M}.  The  success of these
models at explaining  the spread in surface chemical composition
observed in PG 1159 stars \citep{2006A&A...454..845M}, the short
born-again times  of V4334 Sgr \citep{2007MNRAS.380..763M}, and the
location of the GW Vir instability strip in the $\log T_{\rm eff}-\log
g$ plane \citep{2006A&A...458..259C}  renders reliability  to the
inferences drawn for individual pulsating PG 1159 stars. 
It is
worth mentioning, in the context of nonadiabatic analysis of GW Vir
stars, the important theoretical work
of \cite{2004ApJ...610..436Q,2005A&A...441..231Q,2007ApJS..171..219Q,2012ApJ...755..128Q}. These
studies  have shed light on the issues of the presence of variable
and non-variable stars in the GW Vir region of the HR diagram, and the
existence of the high-gravity red edge of the GW Vir instability
domain.

\J0\ is a pulsating PG 1159  
star with $T_{\rm eff}= 90\,000 \pm 900$ K and $\log g= 7.5
\pm 0.01$ (cgs) according to \citet{2006A&A...454..617H}, who employed the 
Data Release 4 of the SDSS Catalogue to estimate these 
quantities using non-LTE (non-local thermodynamic equilibrium) model
atmospheres. \citet{2012MNRAS.426.2137W} performed high-speed photometric 
observations in 2007 and 
2009, and found a set of pulsation frequencies in the range of
$1038$-$3323\ \mu$Hz with amplitudes between $3.5$ and $18.6$ mmag.
The data gathered by \cite{2012MNRAS.426.2137W} shows three
frequencies closely spaced in the 2007 data, which enables, in principle,
to estimate a period rotation for this star. 

\vv\ is  a PNNV star characterized by
$T_{\rm eff}= 130\,000 \pm 13\,000\ $K and $\log g= 7 \pm 0.5\
$ \citep{2006PASP..118..183W}. Its  stellar mass is $M_{\star}= 0.59 \
M_{\sun}$  according to \cite{2006PASP..118..183W} and
$M_{\star}=  0.53 \ M_{\sun}$  according  to the evolutionary tracks  created
by  \citet{2006A&A...454..845M}.  The surface chemical composition
of \vv\ is typical of  PG 1159 stars: C/He= 1.5 and O/He=
0.4 \citep{2006PASP..118..183W}. \vv\  was first observed  as potentially
variable  by \cite{1988PASP..100..187L}.  Later,  it  was monitored
by \cite{1996AJ....111.2332C},  but  no  clear  variability  was  found.
Finally, \cite{2006A&A...454..527G} were able to  confirm 
the ---until then---  elusive intrinsic variability of \vv\  
for the first time. They found  clear evidence that the pulsation  
spectrum of this star is extremely  complex.   Indeed,  the main  
peaks  of the  power
spectrum  have   amplitudes  strongly  variable   between  observation
seasons, and sometimes they are detected only in a particular run.  It
is apparent  that real periodicities of  \vv\ are in  the range $131-
5682$ s.   It is important  to note  that the  shortest periods  in
the observed  period  spectrum  of  \vv\  could  be  associated  with
the $\varepsilon$-mechanism  of mode  driving acting at the 
He burning shell, as suggested by 
\citet{2006A&A...454..527G}. This hypothesis was explored from a theoretical 
point of view by \citet{2009ApJ...701.1008C}. If this hypothesis 
were confirmed, this  object
could  be the first known pulsating PG 1159 star undergoing pulsation
modes powered by this mechanism.

In this work, we present an adiabatic asteroseismological study of 
\J0\ and \vv\
aimed at determining  the internal  structure and  evolutionary status
of these stars on the basis of  the very detailed PG 1159
evolutionary models  of \citet{2006A&A...454..845M}. 
We emphasize that the results presented in this 
work could change to some extent if another
independent set of PG 1159 evolutionary tracks constructed assuming a 
different input physics were employed. We  compute
adiabatic $g$-mode pulsation periods on  PG 1159 evolutionary models
with stellar masses  ranging from $0.515$  to $0.741  \
M_{\sun}$. These  models take into account  the complete evolution of
progenitor  stars, through the thermally pulsing  AGB phase and
born-again episode. A brief summary of the stellar models employed
is provided in Sect. \ref{evolutionary}. We estimate a mean period 
spacing for both \J0\ and \vv\ (Sect. \ref{estimation}),  and then 
we  constrain the stellar  masses
of these stars by comparing the observed period spacing with the
asymptotic period spacing  and with the  average of the computed
period spacings (Sect. \ref{period-spacing}).  In Sect. \ref{fitting}
we employ the individual  observed periods to search for a
representative seismological model for  these  stars.  In
Sect. \ref{splitting} we estimate the rotation period for \J0,
employing the observed triplet of frequencies. Finally,  we  close the
article with a discussion and summary in Sect. \ref{conclusions}.

\begin{figure}  
\centering  
\includegraphics[clip,width=250pt]{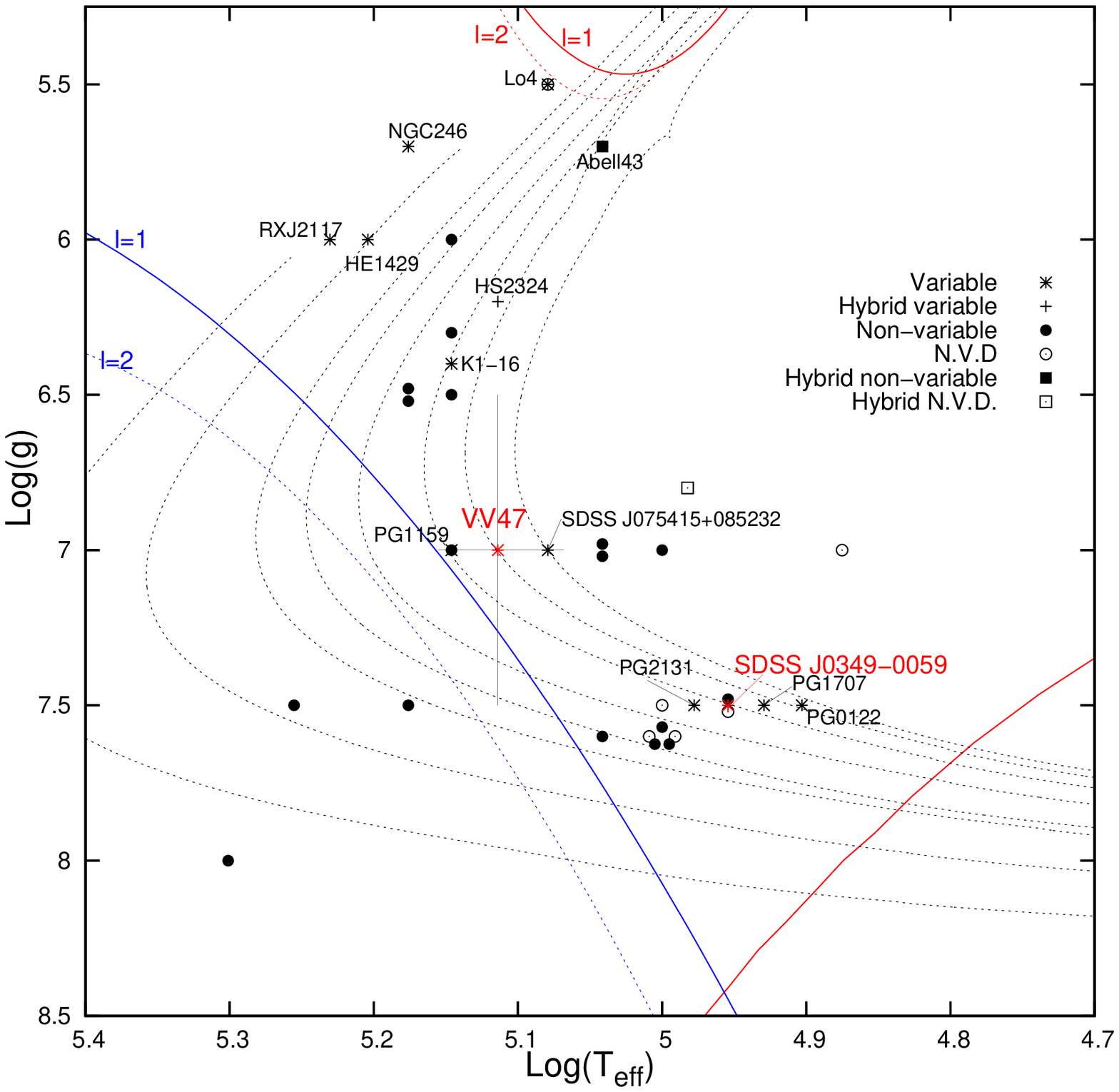}  
\caption{The PG 1159 (VLTP) evolutionary tracks of 
\cite{2006A&A...454..845M} (from right to  left: $M_{\star}= 0.515$,
$0.530$, $ 0.542$,  $0.565$, $0.589$,  $0.609$,  $0.664$,  $0.741 \ M_{\sun}$)
in the $\log T_{\rm eff} - \log g$ diagram (thin dotted curves).  
The location of all known  PG 1159 stars (variable, non-variable, and 
objects with no variability data) is shown,  including \J0 and \vv\ with their 
uncertainties ---the uncertainty values for \J0 are very small in the 
adopted scale and they cannot be seen in the plot. The hot blue edges 
(blue curves) and the low-gravity red edges (upper red curves) 
of  the theoretical GW Vir instability  strip for 
$\ell= 1$ (dashed) and $\ell= 2$ 
(dotted) modes according to \citet{2006A&A...458..259C} 
are also depicted. The high-gravity red edge due to
due to the competition of residual stellar winds 
against the gravitational settling of C and O
\citep{2012ApJ...755..128Q} is also included (lower red solid curve).}
\label{grave-obs}  
\end{figure}  
  
\section{Evolutionary models and numerical tools}  
\label{evolutionary}  
  
The  pulsation  analysis  presented  in  this work  relies  on  a  suite 
of state-of-the-art  stellar  models  that take  into  account the  complete
evolution of the PG 1159 progenitor stars. Specifically, the stellar models
were extracted  from the evolutionary  calculations presented
by  \cite{2005A&A...435..631A}, \cite{2006A&A...454..845M}, and
\cite{2006A&A...458..259C}, who  computed the complete evolution of model
star sequences with initial masses on  the ZAMS in the range $1 - 3.75
\ M_{\sun}$.  All of the  post-AGB evolutionary sequences computed with
the {\tt LPCODE} evolutionary code \citep{2005A&A...435..631A} were followed
through  the  very  late   thermal  pulse  (VLTP)  and  the  resulting
born-again  episode that  give rise  to the  H-deficient, He-,  C- and
O-rich composition characteristic of  PG 1159 stars.  The masses of the
resulting  remnants are  $0.530$, $0.542$, $0.565$, $0.589$, $0.609$,
$0.664$, and  $0.741 \ M_{\sun}$. An additional evolutionary  track of a
remnant with  a stellar mass  of $0.515 \ M_{\sun}$ is  also employed in the 
case of \vv. In Fig.~\ref{grave-obs} the  evolutionary tracks employed  in this 
work are shown in the $\log T_{\rm eff} - \log g$ plane. 
The blue and red edges of the theoretical GW Vir instability domain
according to  \citet{2006A&A...458..259C} correspond to PG 1159 models 
with surface fractional abundances ($X_{\rm i}$) of $^{4}$He in the 
range $0.28-0.48$, $^{12}$C in 
the range $0.27-0.41$ and  $^{16}$O in the range $0.10-0.23$ 
\citep[see Table 1 of][]{2006A&A...458..259C}.
For details about the input physics and evolutionary code used,
and  the   numerical  simulations  performed  to   obtain  the  PG 1159
evolutionary sequences  employed here, we refer  the interested reader
to the works by  \citet{2005A&A...435..631A} and 
\citet{2006A&A...454..845M,2007A&A...470..675M,2007MNRAS.380..763M}

On   the    basis   of   the   evolutionary    tracks   presented   in
Fig.~\ref{grave-obs},  a  value  for the spectroscopic
mass\footnote{According to the nomenclature widely accepted in the 
literature, we use the term ``spectroscopic  mass'', although 
the term  ``evolutionary mass'' might be more appropriate, because
its derivation involves the employment of evolutionary tracks.} of
\J0\ and \vv\ can be derived by linear interpolation. In the case of \J0\
we get a stellar mass of $M_{\star}= 0.543 \pm 0.004 \ M_{\sun}$.   It
is worth mentioning that this is the first time that a mass
determination is done for this star. As for \vv, it is
important to remark that the uncertainties in $\log g$ and $\log
T_{\rm eff}$ are so big than the value derived for the spectroscopic
mass is not very accurate. In this case, we get a formal value
of $M_{\star}\sim 0.529 \ M_{\sun}$ \citep[in agreement  with  the  value  
of $M_{\star}= 0.53\ M_{\sun}$ derived by][]{2006A&A...454..845M}, 
although it can be as low as $\sim 0.510\ M_{\sun}$ or as high as $\sim
0.609\ M_{\sun}$. At this point, it is
worth mentioning the interesting work by \citet{2009JPhCS.172a2077Q},
who proposed a novel approach called ``non-adiabatic asteroseismology''
that leads to a much more precise value of the surface gravity value 
for \vv, of $\log g= 6.1 \pm 0.1$. 
By adopting $T_{\rm eff}= 130\,000$ K and $\log g= 6.1\ $,
our evolutionary tracks predict a spectroscopic mass of $\sim 0.542\
M_{\sun}$.

We  computed $\ell=  1, 2$ $g$-mode  adiabatic pulsation  periods  in
the range  $80-6000$ s  with  the  adiabatic version of the 
{\tt LP-PUL} pulsation code \citep[][]{2006A&A...454..863C}
and the same methods  we employed  in our  previous  works\footnote{La
Plata Stellar Evolution and Pulsation Research Group 
(http://fcaglp.fcaglp.unlp.edu.ar/evolgroup/)}.  We analyzed about 4000
PG 1159 models covering a wide range of effective temperatures
($5.4 \gtrsim \log T_{\rm eff} \gtrsim 4.8$), luminosities
($0 \lesssim   \log(L_*/L_{\sun}) \lesssim 4.2$), and 
stellar masses    ($0.515 \leq M_{\star}/M_{\sun} \leq 0.741$).
 
\section{Estimation of a constant  period spacing}  
\label{estimation}

\begin{table}[ht]
\centering
\caption{List of the 13 independent frequencies in the 2007 January data 
of \J0\ from \cite{2012MNRAS.426.2137W}. The periods that 
more closely 
follow a constant period spacing are emphasized with boldface.}
\begin{tabular}{crcrrrcrrcrc}   %para separar en columnas con lineas, con el pipe aca adentro
\hline
\hline  
$\Pi$ [s] & Freq.[$\mu$Hz] & Ampl. [mmag] \\
\hline  
{\bf 963.48 $\pm 0.37$}& $1037.9 \pm 0.4$ &  $3.7 \pm 0.9$\\
{\bf 906.37 $ \pm 0.33$}& $1103.3 \pm 0.4$ & $3.9 \pm 0.9$\\
$517.84 \pm 0.03$& $1931.1 \pm 0.1$ & $11.3 \pm 1.0$\\
$504.18 \pm 0.05$& $1983.4 \pm 0.2$ &  $6.8 \pm 0.9$\\
{\bf 486.40 $\pm 0.09$}& $2055.9 \pm 0.4$ &  $4.2 \pm 0.9$\\
$482.58 \pm 0.05$& $2072.2 \pm 0.2$ &  $6.6 \pm 0.9$\\
{\bf 465.05 $\pm 0.09$}& $2150.3 \pm 0.4$ & $3.5 \pm 0.9$\\
{\bf 421.48 $\pm 0.04$}& $2372.6 \pm 0.2$ &  $6.7 \pm 0.9$\\
{\bf 418.90 $\pm 0.02$}& $2387.2 \pm 0.1$ &  $18.6 \pm 1.1$\\
{\bf 416.42 $\pm 0.02$}& $2401.4 \pm 0.1$ & $15.2 \pm 1.0$\\
$353.79 \pm 0.03$& $2826.5 \pm 0.2$ & $7.2 \pm 0.9$\\
{\bf 349.02 $\pm 0.01$}& $2865.2 \pm 0.1$ & $12.1 \pm 1.0$\\
{\bf 300.93 $\pm 0.03$}& $3323.0 \pm 0.3$ & $5.5 \pm 0.9$\\
\hline
\hline  
\end{tabular}
\label{table1}
\end{table}

\begin{table}[ht]
\centering
\caption{List of the 10 independent frequencies in the 2009 
December data of \J0\ from \cite{2012MNRAS.426.2137W}. 
The periods that best fit a constant period separation are 
emphasized with boldface.}
\begin{tabular}{crcrrrcrrcrc}
\hline
\hline  
$\Pi$ [s] & Freq.[$\mu$Hz] & Ampl. [mmag] \\
\hline
{\bf 911.49 $\pm 0.08$}& $1097.1 \pm 0.1$ &$9.1 \pm 0.7$\\
{\bf 680.83 $\pm 0.14$}& $1468.8 \pm 0.3$ &  $3.6 \pm 0.7$\\
{\bf 561.83 $\pm 0.09$}& $1779.9 \pm 0.3$ & $4.0 \pm 0.7$\\
$516.72 \pm 0.11$& $1935.3 \pm 0.4$&  $3.3 \pm 0.7$\\
{\bf 511.43 $\pm 0.08$}& $1955.3 \pm 0.3$&  $4.2 \pm 0.7$\\
$419.55 \pm 0.07$& $2383.5 \pm 0.4$& $3.4 \pm 0.7$\\
{\bf 419.18 $\pm 0.02$}& $2385.6 \pm 0.1$&  $15.2 \pm 0.9$\\
$412.27 \pm 0.03$& $2425.6 \pm 0.2$& $10.0 \pm 0.8$\\
{\bf 349.06 $\pm 0.01$}& $2864.8 \pm 0.1$& $9.6 \pm 0.8$\\
$300.93 \pm 0.03$& $3323.0 \pm 0.3$&  $4.3 \pm 0.7$\\
\hline
\hline  
\end{tabular}
\label{table2}
\end{table}

In the asymptotic limit of  high-radial order $k$, nonradial $g$ modes
with the same harmonic degree $\ell$ are expected to be equally spaced 
in period \citep{1980ApJS...43..469T}:

\begin{equation}
\label{eq:spacing}
\Delta\Pi^{\rm a}_{\ell}=\Pi_{k+1,\ell}-\Pi_{k,\ell}=\frac{2\pi^2}{\sqrt{\ell(\ell+1)}} \left[\int^{R_{\star}}_0{\frac{N(r)}{r}dr}\right]^{-1}
\end{equation}

where $N$  is   the  Brunt-V\"ais\"al\"a  frequency.  In principle, 
one can compare the asymptotic period spacing or the average of the 
period spacings computed from a grid of
models with different masses and effective temperatures with
the mean period spacing exhibited by the star, and then infer
the value of the stellar mass (Sect. \ref{period-spacing}). These methods take
full advantage of the fact that the period spacing of pulsating 
PG 1159 stars depends primarily on the stellar mass, and weakly on
the luminosity and the He-rich envelope mass fraction, as it was first 
recognized by \citet{1986PhDT.........2K,1987fbs..conf..297K,
1988IAUS..123..329K,1990ASPC...11..494K}  
\citep[see also][]{1994ApJ...427..415K,2006A&A...454..863C}.
These approaches have been successfully applied in
numerous studies of pulsating PG 1159 stars 
\citep[see, for instance,][]{2007A&A...461.1095C,2007A&A...475..619C,2008A&A...478..869C,2009A&A...499..257C}. The first step in this process is to 
obtain (if exists) a mean uniform
period separation underlying the observed periods. We searched for  a  
constant  period  spacing  in  the data  of the stars under analysis  
by  using  the Kolmogorov-Smirnov  \citep[K-S; see][]{1988IAUS..123..329K}, 
the  Inverse Variance \citep[I-V; see][]{1994MNRAS.270..222O} and
the Fourier Transform \citep[F-T; see][]{1997MNRAS.286..303H} 
significance tests. In the  K-S  test, the
quantity $Q$ is  defined as the probability that  the observed periods
are   randomly  distributed.   Thus,  any   uniform  ---or   at  least
systematically  non-random--- period  spacing present  in  the period
spectrum of the star under analysis will appear as a minimum in $Q$.
In the I-V test, a maximum  of the  inverse  variance  will indicate  
the  presence of  a constant period spacing. Finally, in the F-T test, 
we calculate the Fourier Transform of a Dirac comb 
function (created from a set of observed periods), and then we plot 
the square of the amplitude of the resulting function in terms of  
the inverse of the frequency. And once again, a maximum in the 
square of the amplitude will 
indicate the presence of a constant period spacing.  

\subsection{The case of \J0}
\label{testJ0}

\begin{figure}  
\centering  
\includegraphics[clip,width=250pt]{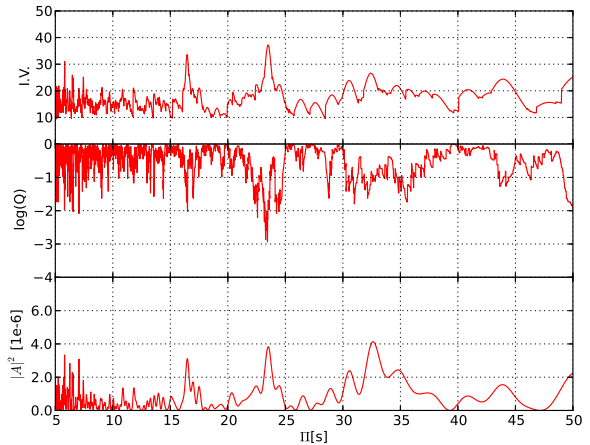}  
\caption{The I-V (upper panel), the K-S (middle panel) and the 
F-T significance (bottom panel) tests applied to the period  spectrum
of \J0 to search for a constant period spacing. The periods  used here
are those indicated in Table~\ref{table1} and \ref{table2}.}
\label{tests-1-jo}  
\end{figure}

\begin{figure}  
\centering  
\includegraphics[clip,width=250pt]{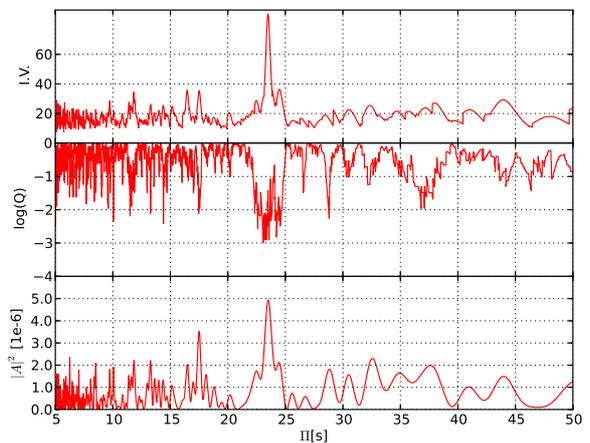}  
\caption{Same as Fig. \ref{tests-1-jo}, but for the 15 selected 
periods of \J0\ emphasized with boldface in Tables \ref{table1} and 
\ref{table2}.}
\label{tests-6-jo}  
\end{figure}

We were able to infer an estimate for the period spacing
of \J0 by using the data available in the work of 
\citet{2012MNRAS.426.2137W}. In particular, we employed the periods 
listed in Tables \ref{table1} and \ref{table2},  extracted from  Table
2 of \citet{2012MNRAS.426.2137W},  corresponding to two  different
observation dates. First, we considered the complete set of periods
from both tables. The results of our analysis are  shown in
Fig.~\ref{tests-1-jo}. The plot shows  evidence of a period spacing at
about $23.5$ s, but it shows  other minima as well, at $\sim 17$ s in
the three tests. Next, we repeated the analysis for several different
sets of data in which we discarded one or two periods,
following \citet{2012MNRAS.426.2137W}. We found an unambiguous
indication of a constant period spacing of $23.49$ s and also  a
secondary solution of $16.5$ s, as it can be seen from
Fig.~\ref{tests-6-jo}, where the periods used are those emphasized
with  boldface in Tables \ref{table1} and \ref{table2}. By comparing
with  the $\ell= 1$ period spacing observed in other GW Vir stars:
$21.6$ s for RXJ2117$+$3412 \citep{2002A&A...381..122V}; $22.9$ s for
PG 0122$+$200 \citep{2007A&A...467..237F}; $21.4$ s for PG
1159$-$035 \citep{2008A&A...477..627C}; $21.6$ s for PG
2131+066 \citep{2000ApJ...545..429R}; $23.0$ s for PG
1707+427 \citep{2004A&A...428..969K}; $22.3$ s for NGC
1501 \citep{1996AJ....112.2699B}; and and also comparing with the
asymptotic period spacing of our PG 1159 models,  we can safely
identify the period spacing of $23.49$ s with $\ell= 1$ modes.
We call the attention that we have discarded the second-highest 
amplitude mode of the 2009 set of modes as well as the fourth-highest 
amplitude mode of the 2007 set of 13 modes (i.e. $\Pi= 412.27\ $s 
and $\Pi= 517.84\ $s, 
respectively) because they do not match with the determined period spacing 
associated with $\ell= 1$. We found that they are probably associated with $\ell=2$ and our models predict their theoretical values to be $410.38\ $s and $518.10\ $s, 
respectively. Regarding the fifth-highest amplitude mode of 
the 2007 set ($\Pi= 353.79\ $s), our models predict that it is associated 
with $\ell= 1$, but its inclusion or exclusion from the list of 
the considered periods does not affect the final result.

\begin{figure}  
\centering  
\includegraphics[clip,width=250pt]{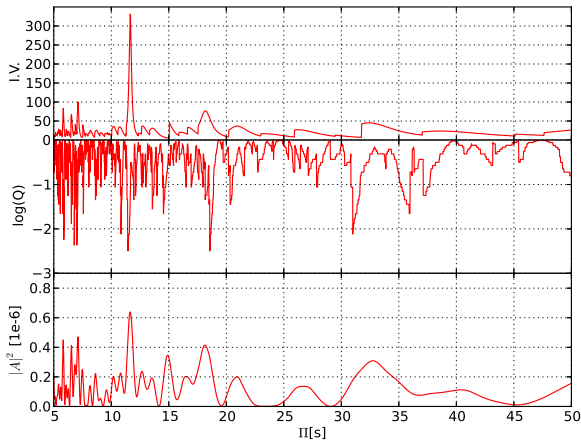}  
\caption{Same as Fig. \ref{tests-1-jo} but for 6 of the discarded 
periods of \J0.}
\label{tests-8-jo}  
\end{figure}

When we apply the three significance tests to 6 of the discarded
periods (i.e. $353.79$, $412.27$, $482.17$, $504.18$, $516.18$ and
$517.84\ $s), we found a period spacing of $11.6$ s, as shown in
Fig.~\ref{tests-8-jo}. At this point, 
we may wonder whether the spacings at  $16.5$ s and $11.6$ s may be 
associated with $\ell= 2$ modes. A simple analysis of the asymptotic period 
spacing  helps us to show that this is not the case. 
Indeed, from Eq. (\ref{eq:spacing}) we can see that
the relation $\Delta\Pi^{\rm a}_{\ell=2}= \Delta\Pi^{\rm a}_{\ell=1}/\sqrt{3}$
between the period spacing corresponding to $\ell= 1$ and $\ell= 2$ holds. 
So, if the period spacing $23.49$ s is associated with $\ell= 1$, then a  
$\Delta\Pi^{\rm a}_{\ell=2}\simeq 13.6$ s should be expected 
for the case of a period separation associated with $\ell= 2$.
From this, we conclude that the spacings of $16.5$ s and $11.6$ s cannot
be associated with $\ell= 2$ modes. 

We refine the value of the period spacing at $23.49$ s by performing 
a least-squares fit to the set of 15 periods employed in building 
Fig.~\ref{tests-6-jo}. We obtain $\Delta \Pi^{\rm O}= 23.4904 \pm 
0.07741$ s. 

In view of the above results, we conclude that there exists strong
evidence for a constant period spacing of $\Delta\Pi^{\rm O}= 23.49\ $s in the
pulsation spectrum exhibited by \J0. 
This is in perfect agreement with \cite{2012MNRAS.426.2137W}. 
We note that the departures from
this period spacing could be associated with the mode-trapping/confinement
phenomena \citep{1994ApJ...427..415K,2006A&A...454..863C}.

\subsection{The case of \vv}
\label{testvv}

The analysis presented in this  Section is based on the work
of \cite{2006A&A...454..527G}. Specifically, we use  the periods
listed in Table \ref{table0}, extracted from Table 4
of \cite{2006A&A...454..527G}, to estimate a period spacing in \vv.
 
\begin{table}  
\centering  
\caption{The most important peaks in the \vv\ power spectra 
according to \cite{2006A&A...454..527G}.  The periods at $240.4$,
$3521$ and $4310\ $s are present with different amplitudes in
different runs. Boldface indicate peaks with best chances to be real.}  
\begin{tabular}{rcc}  
\hline  
\hline  
$\Pi$ [s]& Freq. [$\mu$Hz] &Power [$\mu$mp] \\  
\hline  
131.6         & 7597  & 2.9 \\
132.5         & 7550  & 1.6 \\
153.5         & 6516  & 1.6 \\
163.2         & 6127  & 1.8 \\
189.2         & 5286  & 2.4 \\
211.4         & 4731  & 1.6 \\
240.4         & 4159  & 3.2 \\
              &       & 0.9 \\
{\bf 261.4}   & 3826  & 9.4 \\
280.1         & 3570  & 2.7 \\
1181          &  847  & 1.7 \\ 
1348          &  742  & 1.1 \\
{\bf 2174}    &  460  & 3.6 \\
{\bf 2681}    &  373  & 2.0 \\
2874          &  348  & 1.2 \\
{\bf 3521}    &  284  & 2.1 \\
              &       & 1.2 \\
{\bf 4310}    &  232  & 0.9 \\
              &       & 2.0 \\
              &       & 5.2 \\
5682          &  176  & 1.6 \\
\hline  
\hline  
\end{tabular}  
\label{table0}  
\end{table}  

\begin{figure}  
\centering  
\includegraphics[clip,width=250pt]{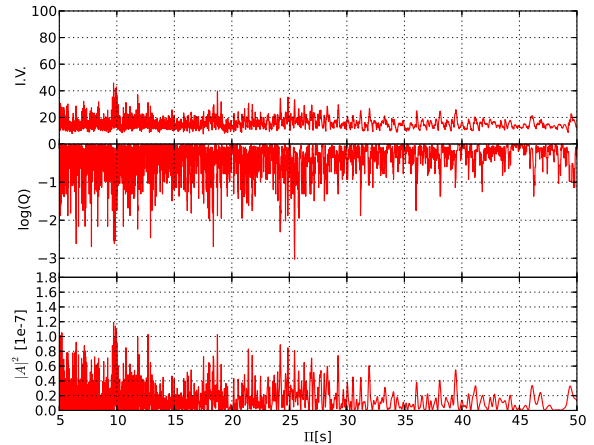}  
\caption{The I-V (upper panel), the K-S (middle panel) 
and the F-T (bottom panel) significance tests applied to the period 
spectrum of \vv\ to search for a constant period spacing. 
The periods used here are those provided by Table \ref{table0}, 
where we considered a single period at $132.05\ $s instead of 
the couple of periods at $131.6$ s and $132.5$ s. 
No unambiguous signal of a constant period spacing is evident.}  
\label{tests-1}  
\end{figure}

First,   we    considered   the    complete   set   of    periods   of
Table \ref{table0}.  For simplicity, we adopted a single period of
132.05 s that  is the average of  the pair of  periods at 131.6 s  and
132.5 s. The results of our analysis  are shown in
Fig.~\ref{tests-1}. In spite of the existence of two minima in $\log
Q$ (middle panel), they are no  statistically meaningful in the
context of the K-S test due to the  presence of other several minima
with  similar significance levels. The F-T based method (lower
panel)  shows multiple local maxima, and then it is not conclusive. As
for the I-V test (upper panel) it does not exhibit any   obvious
maximum. So,  the   application  of  these tests  to  the complete
list  of periods does not  provide us with any  clue about the
existence of a constant period spacing in \vv.

%The results of our analysis are shown in Figs. \ref{tests-1} to \ref{tests-8}.

\begin{figure}  
\centering  
\includegraphics[clip,width=250pt]{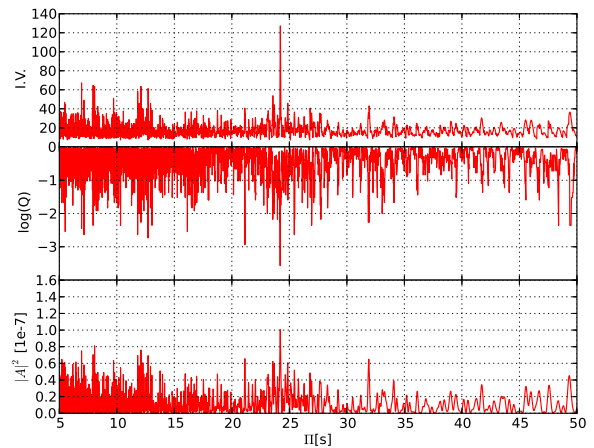}  
\caption{Same as Fig.~\ref{tests-1}, but discarding from the analysis
the periods at 132.05 s, 153.5 s, 240.4 s,  280.1 s, and 3521 s of \vv\
(see text for details).}  
\label{tests-6}  
\end{figure}

Next, we repeated the above analysis for numerous different sets  of
periods, ignoring one or  more periods from the list, and then we
examined the predictions of the three statistical  tests. In
particular,  we found some evidence of the existence of  a period
spacing of  $\sim 24$ s when the periods at $132.05$ s and $153.5$ s
were departed from  the  analysis.  We also  noted  that the  periods
providing  the larger  discrepancies with  that tentative  period
spacing are $240.4$ s, $280.1$ s, $3521$ s , and $4310$ s.
Fig.~\ref{tests-6} is similar to Fig.~\ref{tests-1}   but for  the
situation in  which the  periods at $132.05$ s, $153.5$ s, 
$240.4$ s, $280.1$, s and $3521$ s   are not taken into  account
in the analysis. The plot of the  remaining 11 periods shows strong
evidence of the presence of a constant period spacing at $24.2$ s in
the three tests.

\begin{figure}  
\centering   
\includegraphics[clip,width=250pt]{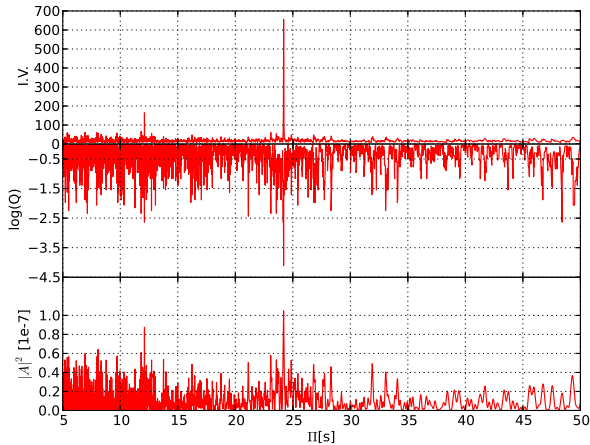}  
\caption{Same as Fig.~\ref{tests-1}, but discarding from the analysis 
the periods at 132.05 s, 153.5 s, 240.4 s, 280.1 s, 3521 s, and  4310 s 
of \vv\ (see text for details).}  
\label{tests-8}  
\end{figure} 

To confirm our hypothesized period spacing, we repeated the three
tests but  this time by  discarding also the  period at 4310  s, which
also presents a significant departure from an equally spaced period
pattern.  We obtained a  very strong confirmation for a period
spacing of $\Delta  \Pi^{\rm O}= 24.2$ s by using  this  subset  of  10
periods.  This   is illustrated in  Fig.~\ref{tests-8}.  The
agreement between   the three methods  is excellent.  The  K-S  test
indicates that   this constant  period spacing is  significant  at  a
confidence level  of $[100\times(1-Q)]= 99 \%$. The F-T also shows
unambiguous evidence of this value.  However, the most clear and
unambiguous indication of $\Delta  \Pi^{\rm O}= 24.2\ $s is provided by the
I-V test. It is also possible to  visualize the first harmonic at
$\Delta \Pi^{\rm O}/2= 12.1$ s shown by all the three tests.  Note that the
value of the maximum in the inverse variance is   more than 5 times
larger than in the case displayed in Fig.~\ref{tests-6}. 

We finally mention that the three tests were applied to the five
periods with best chance to be real according
to \cite{2006A&A...454..527G} (marked with  boldface in
Table \ref{table0}). The results did not  show any clear indication
of the presence of a constant period spacing. 
  
As for \J0, further confirmation for the period spacing 
of \vv\ at $24.2$  s comes from a
least-squares  fit to  the  set of 11  periods  employed in
constructing Fig.~\ref{tests-6}. We obtain $\Delta  \Pi^{\rm O}= 24.2015 \pm
0.03448\ $s. With identical arguments than for \J0\ (see Sect. \ref{testJ0}),
we identify this period spacing with $\ell= 1$ $g$ modes.

In view  of the  above results, we  conclude that there exists evidence for a
constant period spacing  of $\Delta \Pi^{\rm O}= 24.2$ s in the
pulsation spectrum exhibited by \vv, and once again, that the
departures  from this period spacing could  be associated with the
mode-trapping/confinement phenomena as in the case of \J0. It is
important to note that a constant  period spacing at $\sim 24\ $s  is
by far  the largest ever found in  pulsating PG 1159 stars. If we assume
that this period spacing is associated with $\ell= 1$ modes,  a  so high
$\Delta  \Pi^{\rm O}$ value would imply a  rather low total mass  value
for \vv,  in excellent agreement with the  spectroscopic estimation
that  uses  the \cite{2006A&A...454..845M} tracks (see the next
Section).

\section{Mass determination from the observed period spacing}  
\label{period-spacing}  
  
In this section  we constrain the stellar mass of \J0 and \vv\   by
comparing the asymptotic period  spacing and  the average  of the
computed  period  spacings with  the observed period  spacing
estimated in the previous Section for each star.   As mentioned, these
approaches exploit the fact   that the  period  spacing of pulsating
PG 1159 stars  depends  primarily on the stellar mass, and  the
dependence  on  the  luminosity  and the  He-rich  envelope  mass
fraction is  negligible \citep{1994ApJ...427..415K}. In the case 
of \vv, in order to assess  the total mass, we have  
considered both the  high- and low-luminosity regimes of  
the evolutionary sequences, i.e., before and after the 
``evolutionary knee'' of the tracks (see Fig.~\ref{grave-obs}).

We emphasize that the methods to derive the stellar mass
(both spectroscopic and seismic) are not completely independent
because the same set of evolutionary models is used in both approaches
\citep[see][]{2008PASP..120.1043F}. Therefore, an eventual agreement
between spectroscopic and seismic masses only reflects an internal consistency
of the procedure.
  
\subsection{First method: comparing the observed period   
spacing ($\Delta \Pi^{\rm O}_{\ell}$) with the asymptotic period
spacing ($\Delta  \Pi_{\ell}^{\rm  a}$)}    
\label{sect-aps}  
    
\begin{figure}  
\centering  
\includegraphics[clip,width=250pt]{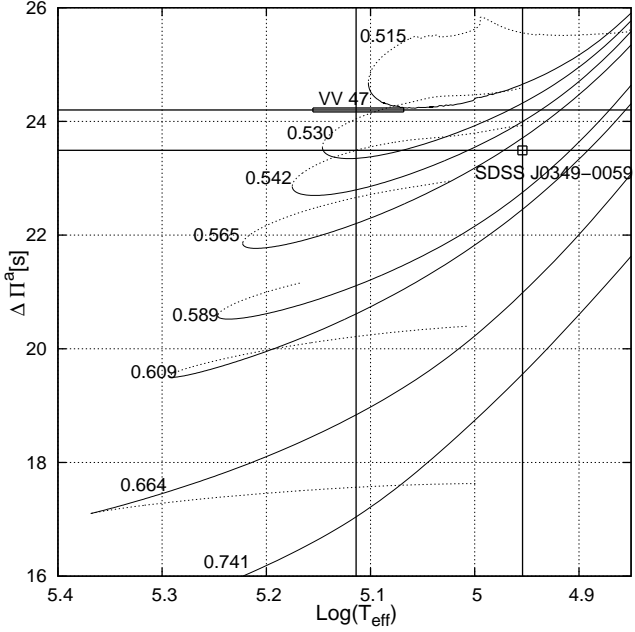}  
\caption{The dipole ($\ell= 1$) asymptotic period spacing, 
$\Delta  \Pi_{\ell}^{\rm  a}$, corresponding 
to each PG 1159 evolutionary sequence considered in this work
(dashed and solid curves), in terms of  the  logarithm 
of the effective temperature.  Numbers  along each  curve denote  
the  stellar mass
(in   solar  units). Dashed (solid)  lines  correspond to evolutionary
stages  before (after)  the turning point at  the maxima  effective
temperature  of each track (``the evolutionary knee'').
The observed period spacing, $\Delta \Pi_{\ell= 1}^{\rm O}$, 
derived for \J0\ and \vv\ are depicted with 
horizontal solid lines. The uncertainties in  
$\Delta \Pi_{\ell= 1}^{\rm O}$ and $T_{\rm eff}$ for each star 
are also indicated.}
\label{asintvvj0}  
\end{figure}

Fig.~\ref{asintvvj0} shows the  asymptotic period spacing
$\Delta \Pi_{\ell}^{\rm a}$ for $\ell= 1$  modes (calculated according
to  Eq.~\ref{eq:spacing}) as a  function  of the  effective
temperature  for different stellar masses.  Also shown in this
diagram are the  locations of \J0, with $\Delta \Pi^{\rm O}_{\ell=
1}= 23.4904 \pm 0.07741\ $s (Sect. \ref{testJ0}) and $T_{\rm eff}=
90\,000 \pm 900\ $K (H\"ugelmeyer et al. 2006), along with \vv, with
$\Delta  \Pi^{\rm O}_{\ell=  1}= 24.2015 \pm 0.03448\ $s
(Sect. \ref{testvv}) and $T_{\rm eff}= 130\,000\ \pm 13\,000$K
(Werner \& Herwig 2006). As can be seen from the Figure, the greater the 
stellar mass, the smaller the values of the asymptotic 
period spacing.

By performing a linear interpolation of the 
theoretical values of $\Delta \Pi_{\ell}^{\rm  a}$, 
the comparison between $\Delta \Pi^{\rm O}_{\ell= 1}$  and
$\Delta  \Pi_{\ell=1}^{\rm  a}$ yields  a  stellar mass  of
$M_{\star}= 0.569^{+0.004}_{-0.002}\ M_{\sun}$ for \J0.
Proceeding similarly for \vv\ ---when there were
no points to perform the linear interpolation, we extrapolated the 
theoretical values of $\Delta \Pi_{\ell}^{\rm  a}$--- the comparison between
$\Delta \Pi^{\rm O}_{\ell= 1}$ and $\Delta  \Pi_{\ell=  1}^{\rm  a}$
gives  a  stellar mass  of $M_{\star}= 0.520 ^{+0.002}_{-0.005} \ M_{\sun}$
($M_{\star}= 0.526 ^{+0.007}_{-0.005} \ M_{\sun}$)
if \vv\ is after  (before)  the  evolutionary  knee.
Both  values  are  in excellent agreement each other. 
The errors quoted for \vv\ are admittedly tiny and they are 
unrealistic because of the large uncertainties in the effective 
temperature. So, they only represent the internal errors involved in 
the procedure.

As in our previous works, we emphasize that the derivation of the
stellar mass using  the asymptotic period spacing may  not be entirely
reliable in pulsating PG 1159 stars  that exhibit short and
intermediate periods \citep[see][]{2008A&A...478..175A}. This is
because the asymptotic predictions are  strictly valid in the limit of
very high  radial order (very long periods) and  for chemically
homogeneous stellar  models, while  PG 1159  stars are  supposed  to
be  chemically stratified  and characterized  by strong  chemical
gradients  built up during the progenitor star's life.

In the next Section we employ another method to infer the stellar mass
of PG 1159 stars which, even though it is computationally  expensive
because requires detailed  pulsation computations, it has the
advantage of being more realistic. 
  
\subsection{Second method: comparing the observed period   
spacing ($\Delta \Pi^{\rm O}_{\ell}$) with the average of the computed   
period spacings ($\overline{\Delta \Pi_{\ell}}$)}    
\label{sect-psp}  

\begin{figure}  
\centering  
\includegraphics[clip,width=250pt]{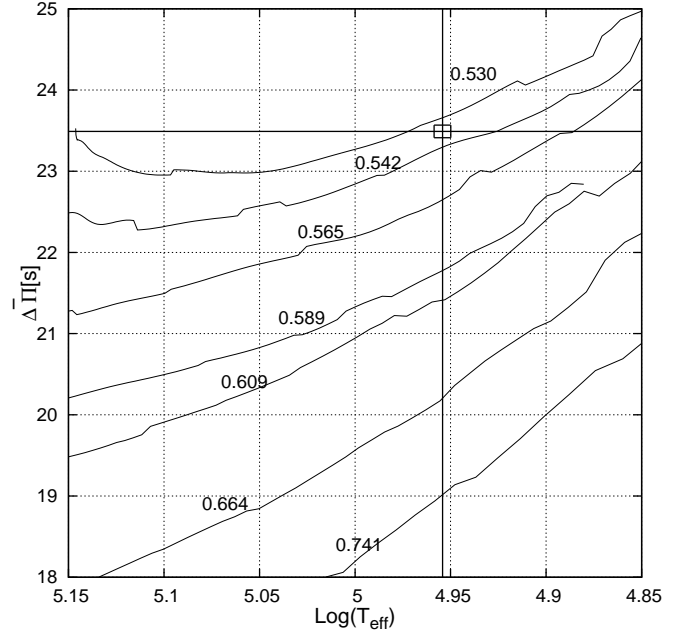}  
\caption{The dipole  average of the computed   
period spacings, $\overline{\Delta \Pi_{\ell}}$, computed 
in the range of periods observed in \J0, corresponding
to each PG 1159 evolutionary sequence considered in this work
(solid curves), in terms of  the  logarithm 
of the effective temperature.  Numbers  along each  curve denote  
the  stellar mass (in   solar  units). 
The observed period spacing, $\Delta \Pi_{\ell= 1}^{\rm O}$, derived 
for \J0\ is depicted with a
horizontal solid line. The uncertainties in  
$\Delta \Pi_{\ell= 1}^{\rm O}$ and $T_{\rm eff}$ are also indicated.}
\label{promj0}  
\end{figure}  

The  average   of  the  computed   period  spacings  is   assessed  as
$\overline{\Delta  \Pi_{\ell}}= (n-1)^{-1}  \sum_{k}  \Delta \Pi_k  $,
where the ``forward'' period spacing is defined as
$\Delta \Pi_k= \Pi_{k+1}-\Pi_k$ ($k$ being the radial order) and $n$
is the number of  theoretical  periods within the  range of  periods
observed in the target star. For \J0, $\Pi_k \in [300,970]\ $s and 
for \vv, $\Pi_k \in [160,5700]\ $s.
  
In  Fig.~\ref{promj0} we  show the  run of  the average of  the
computed period  spacings  ($\ell= 1$)  for  \J0\  in  terms of  the
effective temperature for our PG 1159 evolutionary sequences (without
the sequence of $M_{\star}= 0.515 \ M_{\sun}$), along with 
the observed period spacing for \J0.  As can be appreciated from
the Figure, the greater the 
stellar mass, the smaller the values of the average of  the
computed period  spacings. By performing a
linear interpolation of the theoretical values of
$\overline{\Delta \Pi_{\ell}}$, the comparison between 
$\Delta \Pi^{\rm O}_{\ell= 1}$  and
$\overline{\Delta  \Pi_{\ell=1}}$ yields a stellar  mass  of
$M_{\star}= 0.535 \pm 0.004 \ M_{\sun}$.  This value is  smaller than
the derived through the asymptotic period spacing, showing that the
asymptotic  approach overestimates the  stellar mass of PG 1159 stars
in the case of stars pulsating with short and intermediate periods, 
which is precisely the situation of \J0.  The value obtained by
this approach is more reliable because the method is valid for short,
intermediate and long periods, as long as the average of the computed
period spacing is evaluated at the right range of periods.

In order to investigate the possibility that the  period
spacings at $\sim 16.5$ and $\sim 11.6$ s found  in Sect. \ref{testJ0}
could be associated with quadrupole ($\ell= 2$) modes, we repeated these
calculations, but this time for $\ell= 2$.  If $\Delta \Pi^{\rm
O}_{\ell= 2} \sim 16.5$ s, then the stellar mass  should be much lower
than  $M_{\star}= 0.530\ M_{\sun}$. On the other hand, if
$\Delta \Pi^{\rm O}_{\ell= 2} \sim 11.6$ s,  then the stellar mass
should be $\sim 0.741\ M_{\sun}$. In conclusion,  both possible period
spacing would lead to values  of the stellar  mass too different than
the ones given by the other determinations, in particular,  the one
derived using the spectroscopic parameters 
($M_{\star}= 0.543\pm 0.004 M_{\sun}$, Sect. \ref{evolutionary}). This analysis, 
along with the arguments presented in Sect. \ref{testJ0} based on 
the relation between the asymptotic period spacings for $\ell= 1$ and 
$\ell= 2$, allows to definitely  discard  these values as probable 
period spacings associated with $\ell= 2$.

\begin{figure}  
\centering  
\includegraphics[clip,width=250pt]{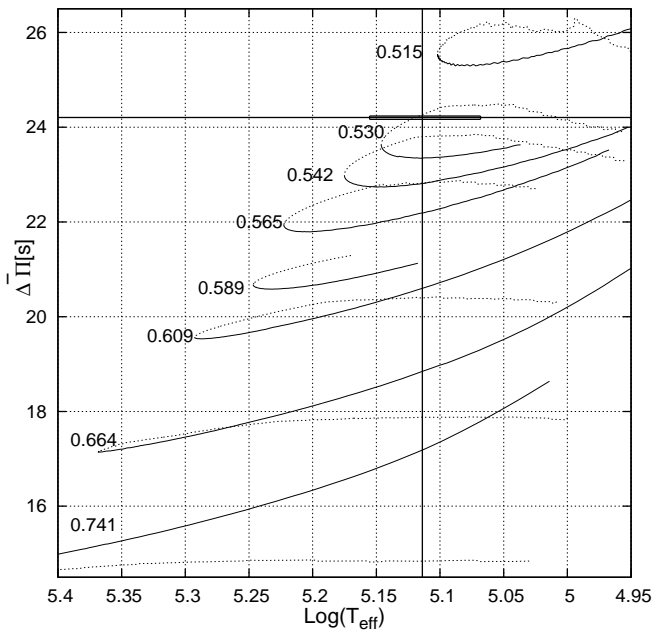}  
\caption{Same as Fig.~\ref{promj0} but for the case of \vv.}
\label{promvv}  
\end{figure}

In  the case of \vv, we depict in Fig.~\ref{promvv} the  run of
average of  the computed period  spacings ($\ell= 1$)  in  terms of
the  effective temperature for our PG 1159 evolutionary sequences
(including $M_{\star}= 0.515 \ M_{\sun}$). The observed period spacing 
for \vv\ is also shown. By adopting the effective
temperature of \vv\  as given by spectroscopy, we found a stellar  mass
of   $M_{\star}= 0.532^{0.004}_{0.007} \ M_{\sun}$  if  the  star   is
``before  the knee'', and  $M_{\star}= 0.524^{0.002}_{0.001}  \
M_{\sun}$  if the  star is  ``after the knee''. 
These values are very close to those derived from the asymptotic
period spacing, as it would be  expected on the grounds that
this star is pulsating with very long periods, almost in the 
asymptotic regime. 

\section{Constraints from the individual observed periods}  
\label{fitting}  

In this procedure we search for a pulsation model that best matches
the
\emph{individual} pulsation periods of a given star under study. The  
goodness of  the  match  between the   theoretical pulsation  periods
($\Pi_k^{\rm  T}$) and  the observed   individual  periods
($\Pi_i^{\rm  O}$) is measured by means  of a merit function defined
as 

\begin{equation}
\label{chi}
\chi^2(M_{\star},  T_{\rm   eff})=   \frac{1}{m} \sum_{i=1}^{m}   \min[(\Pi_i^{\rm   O}-   \Pi_k^{\rm  T})^2], 
\end{equation}

\noindent where $m$ is the number of observed periods. The PG 1159 model that
shows the lowest value of $\chi^2$, if exists, is adopted as the ``best-fit
model''.  This approach has been used in GW Vir stars by  
\citet{2007A&A...461.1095C,2007A&A...475..619C,2008A&A...478..869C,
2009A&A...499..257C} and \citet{2014MNRAS.442.2278K}. We assess the function
$\chi^2=\chi^2(M_{\star}, T_{\rm eff})$ for stellar masses of $0.515$,
$0.530$, $0.542$, $0.565$, $0.589$, $0.609$, $0.664$, and $0.741 \
M_{\sun}$.  For the effective temperature we employ a much finer grid
($\Delta T_{\rm eff}= 10-30$ K) which is given by the time step
adopted by our evolutionary calculations.

\subsection{Searching for the best-fit model for \J0}  
\label{searchingj0}  
  
\begin{figure*}[ht!]
  \begin{center} \subfigure[$\ell= 1$, 10periods]
   {\label{fig:l1ajustej0}
    \includegraphics[width=0.33\textwidth]{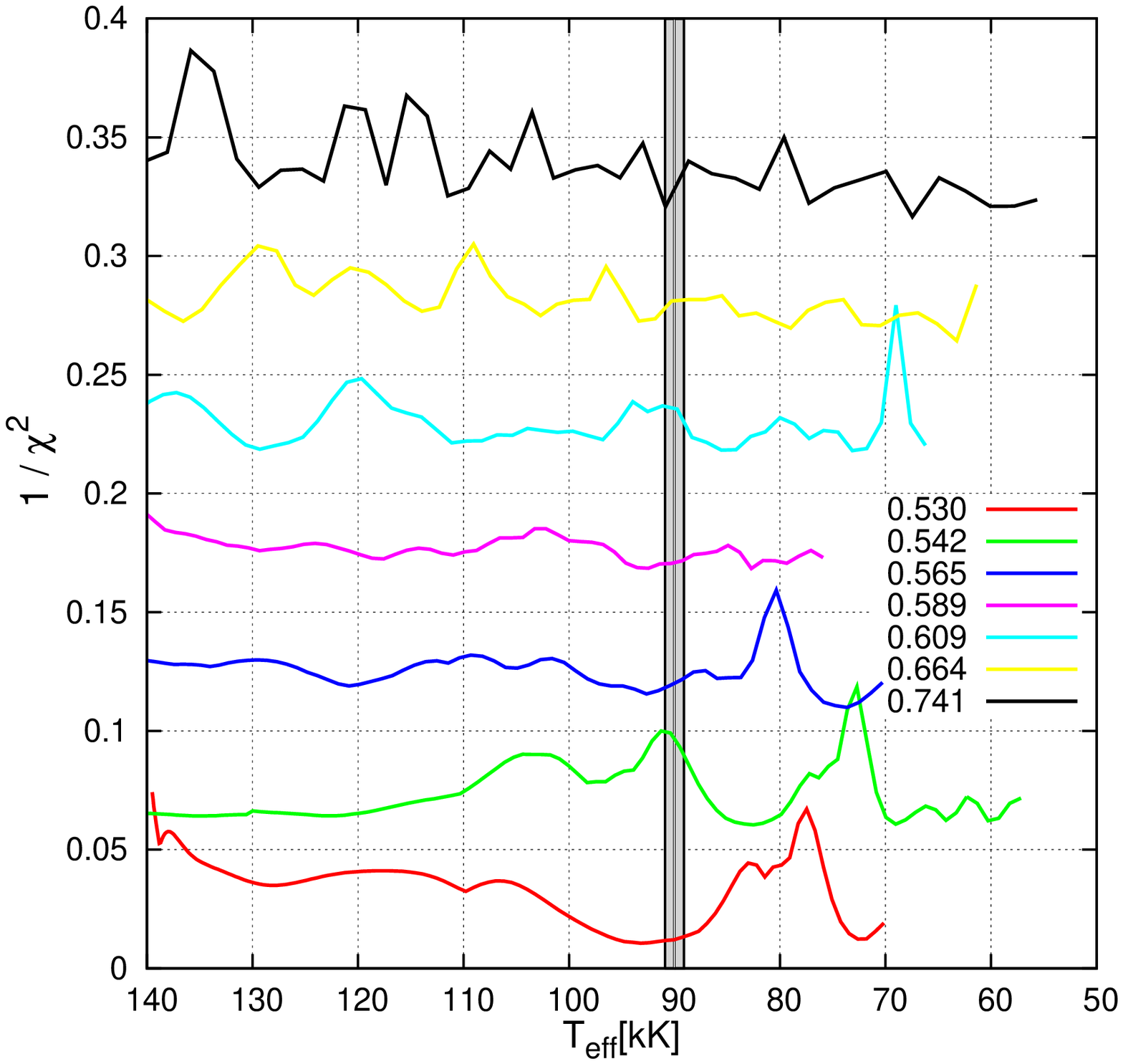}} 
    \subfigure[$\ell= 1$, 11 periods]
     {\label{fig:l1segundoajustej0}
     \includegraphics[width=0.33\textwidth]{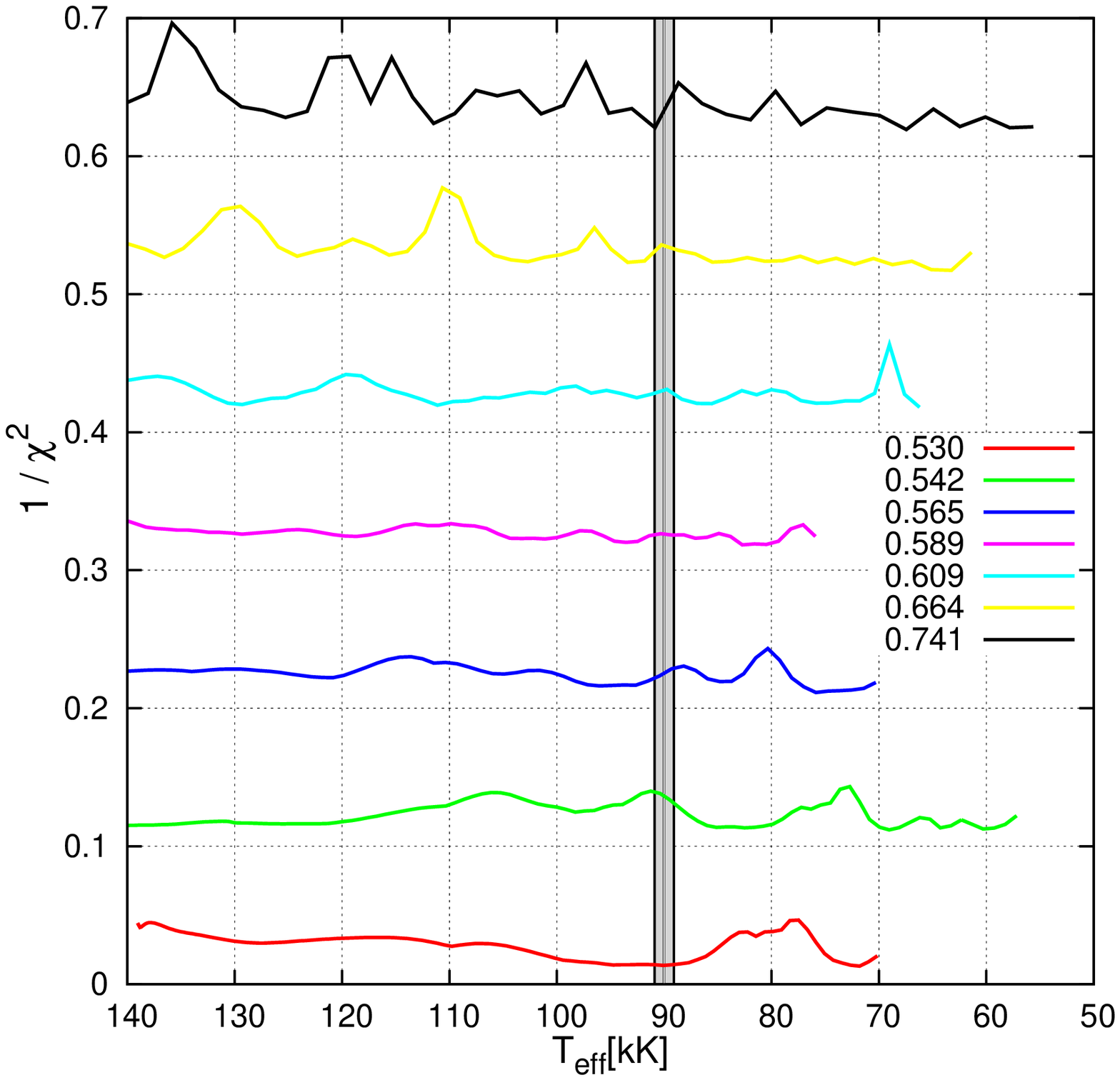}} 
     \end{center} 
    \caption{The inverse of the quality function of the period fit considering
    $\ell= 1$ versus $T_{\rm  eff}$ for \J0, adopting 
    the sets of 10 (left panel) and 11 (right panel) periods 
    (see text for details).      The vertical gray strip
    depicts  the spectroscopic $T_{\rm  eff}$ of \J0\ and its
    uncertainties. For clarity, the curves have been arbitrarily
    shifted upward (with  a step of 0.05 in the left panel and 0.1 in the
    right panel) except for the lowest curve. The adopted model
    as the ``asteroseismological model''  has     
    $M_{\star}= 0.542\ M_{\sun}$. [Color figure
    only available in the electronic version of the article].}  
\label{fig:ajustej0}
\end{figure*}

We start our analysis assuming that  all of  the observed
periods correspond  to $\ell=  1$ modes and consider two different set
of observed periods  $\Pi_i^{\rm  O}$ to compute the quality function
given by Eq.~(\ref{chi}).  We begin by considering the set of periods 
emphasized in boldface in Tables~\ref{table1} and ~\ref{table2} but with a
difference: the values numerically too close each other 
were averaged ---it is not expected that they correspond to independent 
modes--- resulting at the end in a total of 10 periods to be included 
in the period fit. Fig.~\ref{fig:l1ajustej0} shows the quantity
$(\chi^2)^{-1}$ in terms of the effective temperature  for different
stellar  masses, taking into account this set of periods. We also
include the effective temperature of  \J0\ and its uncertainty (vertical 
lines). As mentioned before, the goodness of the match between the 
theoretical and the observed periods is measured by the value of $\chi^2$: 
the better the period match, the lower the value of $\chi^2$ 
---in our case, the greater the value of $(\chi^2)^{-1}$.  For this case,
there is a strong maximum of $(\chi^2)^{-1}$ for a model with
$M_{\star}= 0.609  \ M_{\sun}$ and  $T_{\rm  eff} \sim 69\, 000\ $K as
it can be seen in Fig.~\ref{fig:l1ajustej0}. But, despite of representing
a very good agreement between the observed and the theoretical periods, 
the effective temperature of this possible solution is unacceptably far 
from the $T_{\rm eff}$ of \J0. Another less pronounced maximum correspond
to models with: $M_{\star}= 0.542  \ M_{\sun}$ and 
$T_{\rm  eff} \sim 72\, 000$ K, $M_{\star}= 0.565  \ M_{\sun}$ and 
$T_{\rm  eff} \sim 80\, 000$ K, $M_{\star}= 0.530  \ M_{\sun}$ and $T_{\rm
eff} \sim 77\, 000$ K, and finally,  $M_{\star}= 0.542  \
M_{\sun}$ and $T_{\rm  eff} \sim 91\, 000$ K. This last model is
closer to the effective temperature of \J0\ and thus, we adopt 
this model as the ``asteroseismological model''.

Next, we carried out a period fit considering a set of periods from
Tables~\ref{table1} and ~\ref{table2}, but adding a set of 6 periods
previously discarded (i.e. $353.79$ s, $412.27$ s, $482.58$ s, $504.18$ s,
$516.72$ s, and $517.84$ s). Once again, the values numerically too
close each other were averaged, so it resulted finally in a set of 11 
periods. The
outcome of the period fit in this case is displayed in
Fig.~\ref{fig:l1segundoajustej0}. As compared with the results from the
first period fit (Fig.~\ref{fig:l1ajustej0}), the function
$(\chi^2)^{-1}$ is characterized by a smoother behavior, although the
peak at $\sim 91\,000$ K still remains. As a result, there are not
new solutions.

For the adopted asteroseismological model, we can compare the observed
and the theoretical periods ($\ell= 1$, $m= 0$) by computing the
absolute period differences $|\delta\Pi|= |\Pi^{\rm O}-\Pi^{\rm T}|$. 
The results are shown in Table \ref{tab:tablaper1}. 
Column 5 of Table \ref{tab:tablaper1} shows the value of the 
linear nonadiabatic growth rate ($\eta$), defined as
$\eta$ ($\equiv -\Im(\sigma)/ \Re(\sigma)$,  where $\Re(\sigma)$ 
and $\Im(\sigma)$ are the real and the  imaginary part, respectively, 
of the complex eigenfrequency  $\sigma$, computed 
with the nonadiabatic version of the {\tt LP-PUL} pulsation code 
\citep{2006A&A...458..259C}. A value of $\eta > 0$ ($\eta < 0$) 
implies an unstable (stable) mode (see column 6 of 
Table \ref{tab:tablaper1}). Interestingly enough, 
most of the periods of the asteroseismological 
model for SDSS J0349$-$0059 are associated  to unstable modes. Our 
nonadiabatic computations fails to predict the existence of the unstable 
modes with periods at $\sim 909$ s and $\sim 963$ s observed in the star.
The main features
of the adopted asteroseismological model for \J0 are summarized 
in Table \ref{tab:modeloj0}.

\begin{table}[ht]
\centering
\caption{Observed periods compared with theoretical periods 
corresponding to the the adopted asteroseismological 
model for \J0, with $M_{\star}= 0.542\ M_{\sun}$ and 
$T_{\rm eff}= 91\,255$ K ($\ell= 1, m= 0$). 
Also shown are the radial order $k$, the absolute period difference,
and the nonadiabatic growth rate for each theoretical 
period.}
\begin{tabular}{cccccc}
\hline
\hline
 $\Pi^{\rm O}$[s] & $\Pi^{\rm T}$[s] & $k$ & $|\delta\Pi|$[s] & $\eta$ & 
Remark\\
\hline
$300.93$& $306.45$&  $11$&  $5.52$ & $9.02 \times 10^{-8}$ & unstable \\
$349.04$& $354.92$&  $13$&  $5.88$ & $6.48 \times 10^{-7}$ & unstable \\
$419.04$& $422.45$&  $16$&  $3.41$ & $4.05 \times 10^{-6}$ & unstable \\
$465.05$& $468.23$&  $18$&  $3.18$ & $7.60 \times 10^{-6}$ & unstable \\
$486.40$& $490.69$&  $19$&  $4.29$ & $1.19 \times 10^{-5}$ & unstable \\
$511.43$& $514.65$&  $20$&  $3.22$ & $9.50 \times 10^{-6}$ & unstable \\
$561.83$& $560.07$&  $22$&  $1.76$ & $2.11 \times 10^{-5}$ & unstable \\
$680.83$& $674.99$&  $27$&  $5.84$ & $1.74 \times 10^{-5}$ & unstable \\
$908.93$& $909.39$&  $37$&  $0.46$ & $-2.79 \times 10^{-4}$ & stable \\
$963.48$& $957.00$&  $39$&  $6.48$ & $-7.38 \times 10^{-4}$ & stable \\
\hline
\hline
\end{tabular}
\label{tab:tablaper1}
\end{table}

\begin{table}[ht]
\centering
\caption{Main characteristics of the adopted 
asteroseismological model for \J0.}
\begin{tabular}{c|c|cc}
\hline
\hline
Quantity & Spectroscopy & Asteroseismology & \\
\hline
 $T_{\rm eff}$ [K]  &     $90\,000 \pm 900$ &  $91\,255$&  \\
 $\log(g)$          &     $7.5 \pm 0.01$&  $7.488$&  \\
 $M_{\star}[M_{\sun}]$&     --- &            $0.542$&  \\
 $\log(R_*/R_{\sun})$&     --- &             $-1.658$&  \\
 $\log(L_*/L_{\sun})$&     --- &             $1.475$&  \\
\hline
\hline
\end{tabular}
\label{tab:modeloj0}
\end{table}

\begin{figure*}[ht!]
  \begin{center}
    \subfigure[$\ell= 1, 2$, 10 periods]{\label{fig:mezclaprimerajustej0}\includegraphics[width=0.33\textwidth]{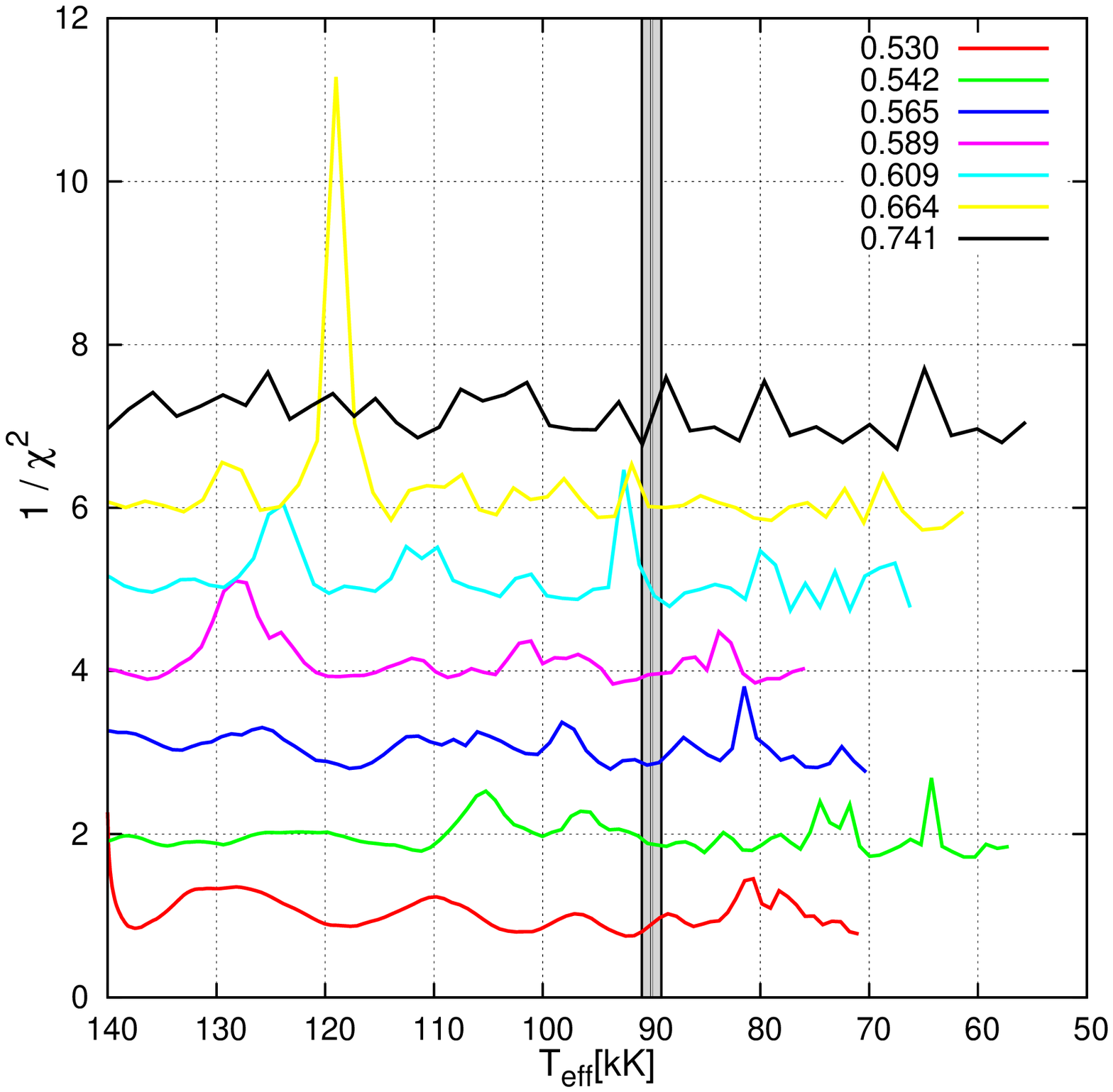}} 
    \subfigure[$\ell= 1, 2$, 11 periods]{\label{fig:mezclasegundoajustej0}\includegraphics[width=0.33\textwidth]{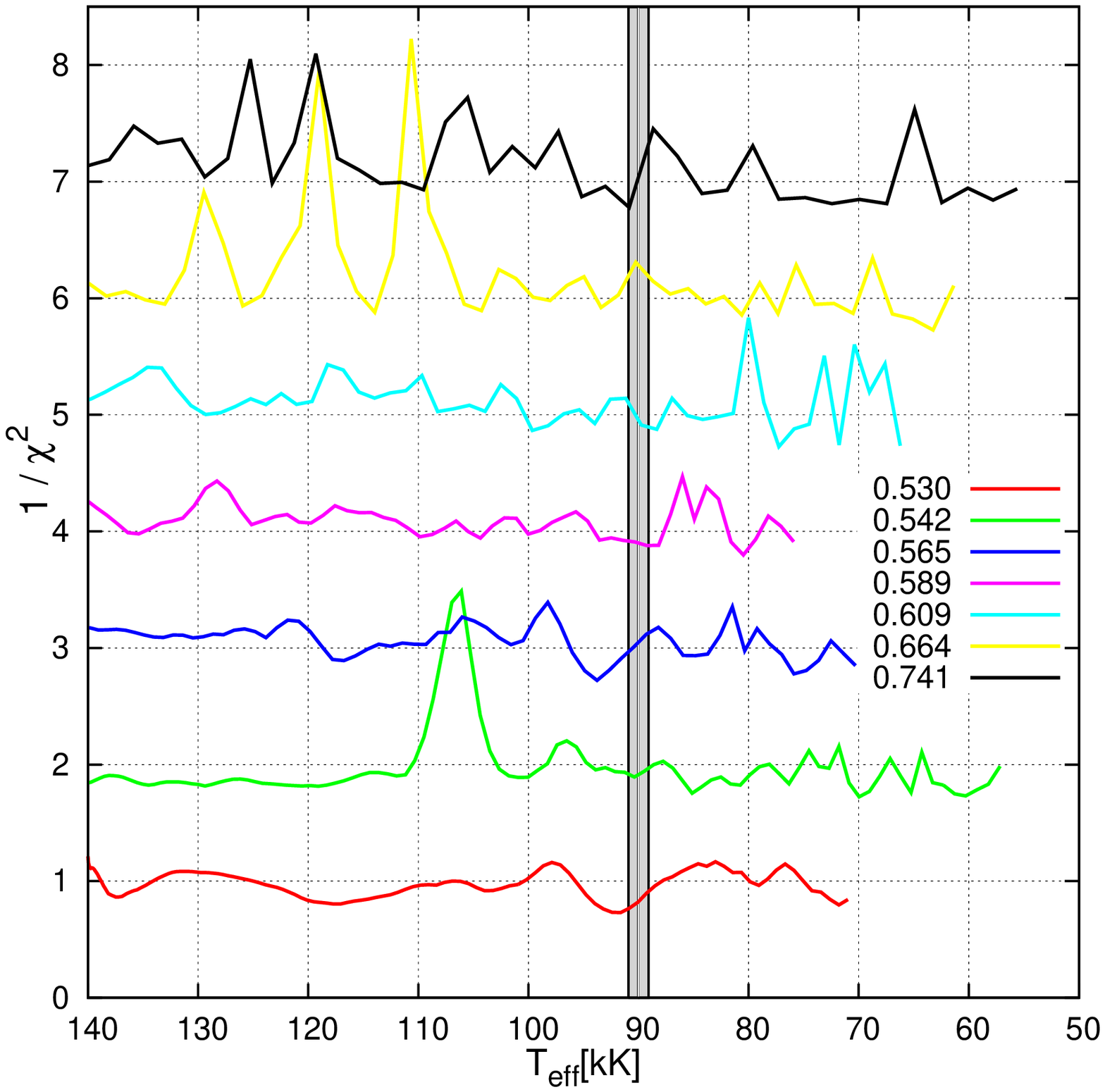}} 
  \end{center}
  \caption{Same as Fig.~\ref{fig:ajustej0}, but comparing the 
observed periods with theoretical $\ell= 1$ and $\ell= 2$ periods. 
In both panels the curves have been arbitrarily shifted upward 
(with a step of 1) except for the lower curve. 
[Color figure only available in the electronic version of the article].}  
\label{fig:mezclaajustej0}
\end{figure*}

Next, we considered the possibility that the periods exhibited by the star
is actually a mix of $\ell= 1, 2$ $g$ modes.  The results are shown in
Fig.~\ref{fig:mezclaajustej0} for the same set of observed periods 
considered in Fig. \ref{fig:l1ajustej0}. In the first
case (Fig.~\ref{fig:mezclaprimerajustej0}) there are new possible
solutions which in principle can represent good period fits, 
but the peaks lie well beyond the range allowed by the uncertainties in 
the effective temperature of \J0. As for the second set of 11 periods 
displayed in
Fig.~\ref{fig:mezclasegundoajustej0}, there are new possible solutions
closer to the effective temperature of \J0, but
there is not a definite asteroseismological model. Altogether, 
since the period fits considering a mix of $\ell= 1$ and $\ell= 2$ modes
do not show a clear solution,  the
results point out into the conclusion that the modes of \J0\
 may probably be associated with $\ell= 1$ only. 

\subsection{Searching for the best-fit model for \vv}  
\label{searchingvv} 
   
\begin{figure*}[ht!]
  \begin{center} \subfigure[16
    periods]{\label{fig:antes-1}\includegraphics[width=0.29\textwidth]{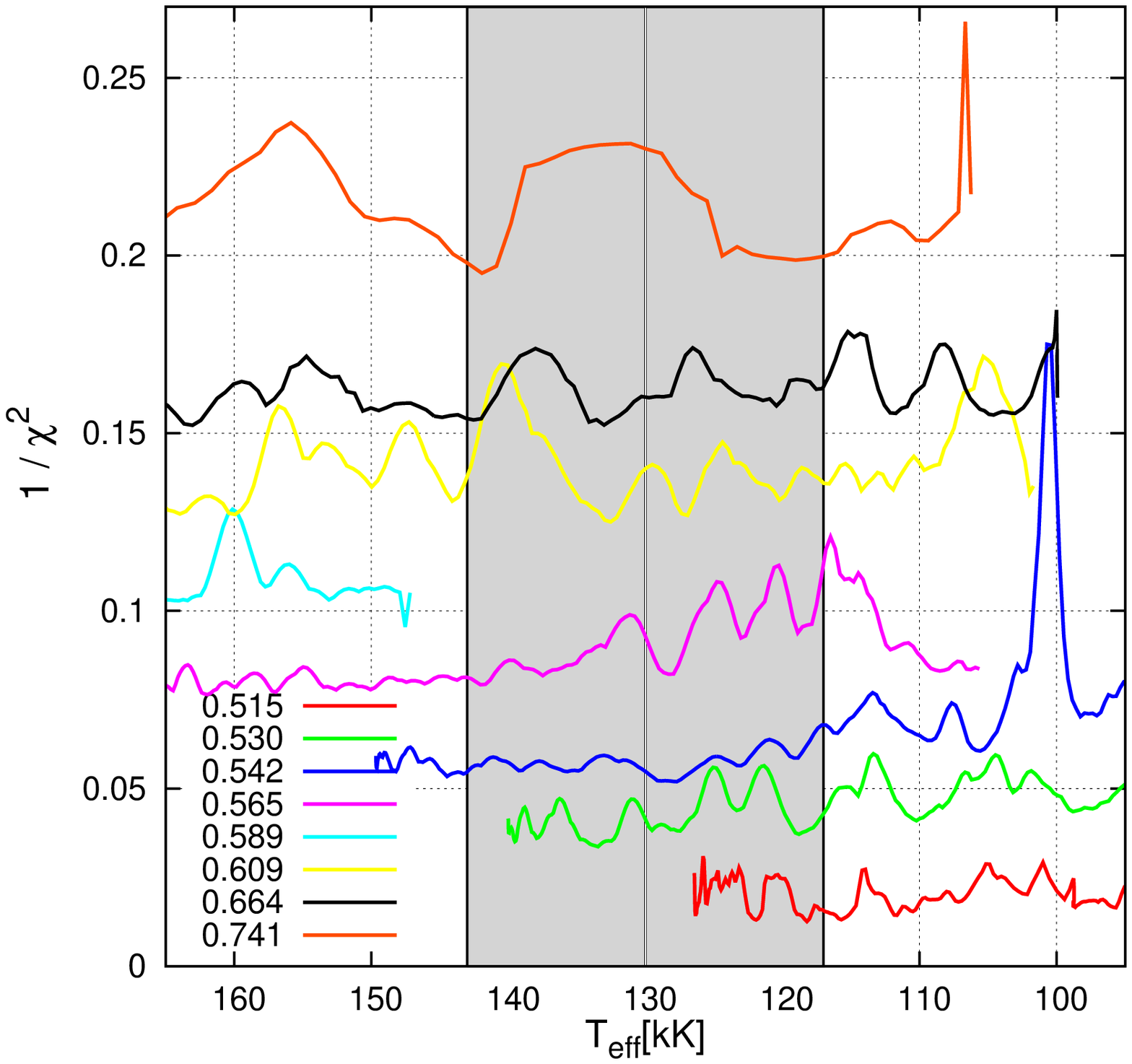}} \subfigure[11
    periods]{\label{fig:antes-3}\includegraphics[width=0.29\textwidth]{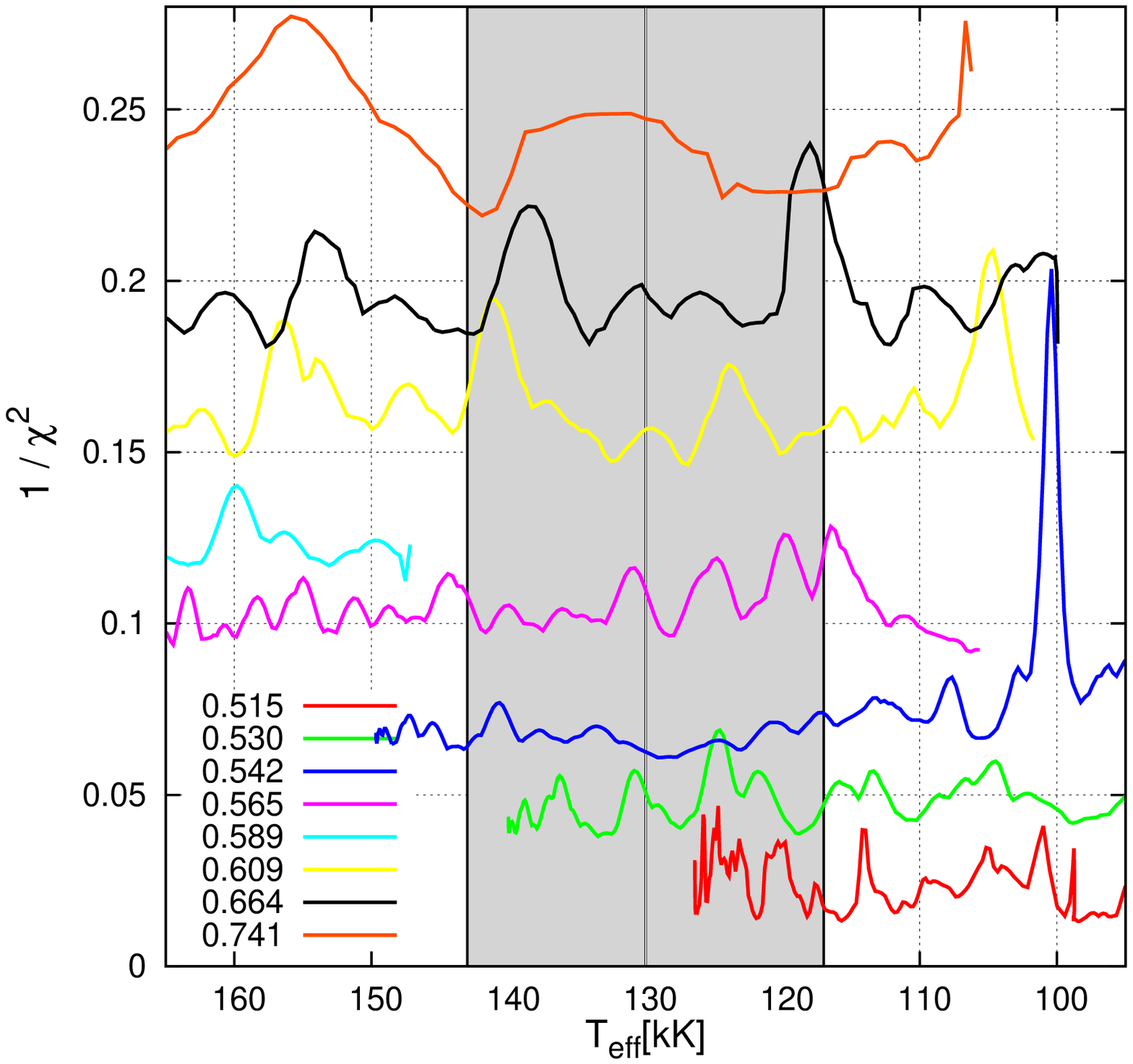}}  \subfigure[8
    periods]{\label{fig:antes-5}\includegraphics[width=0.29\textwidth]{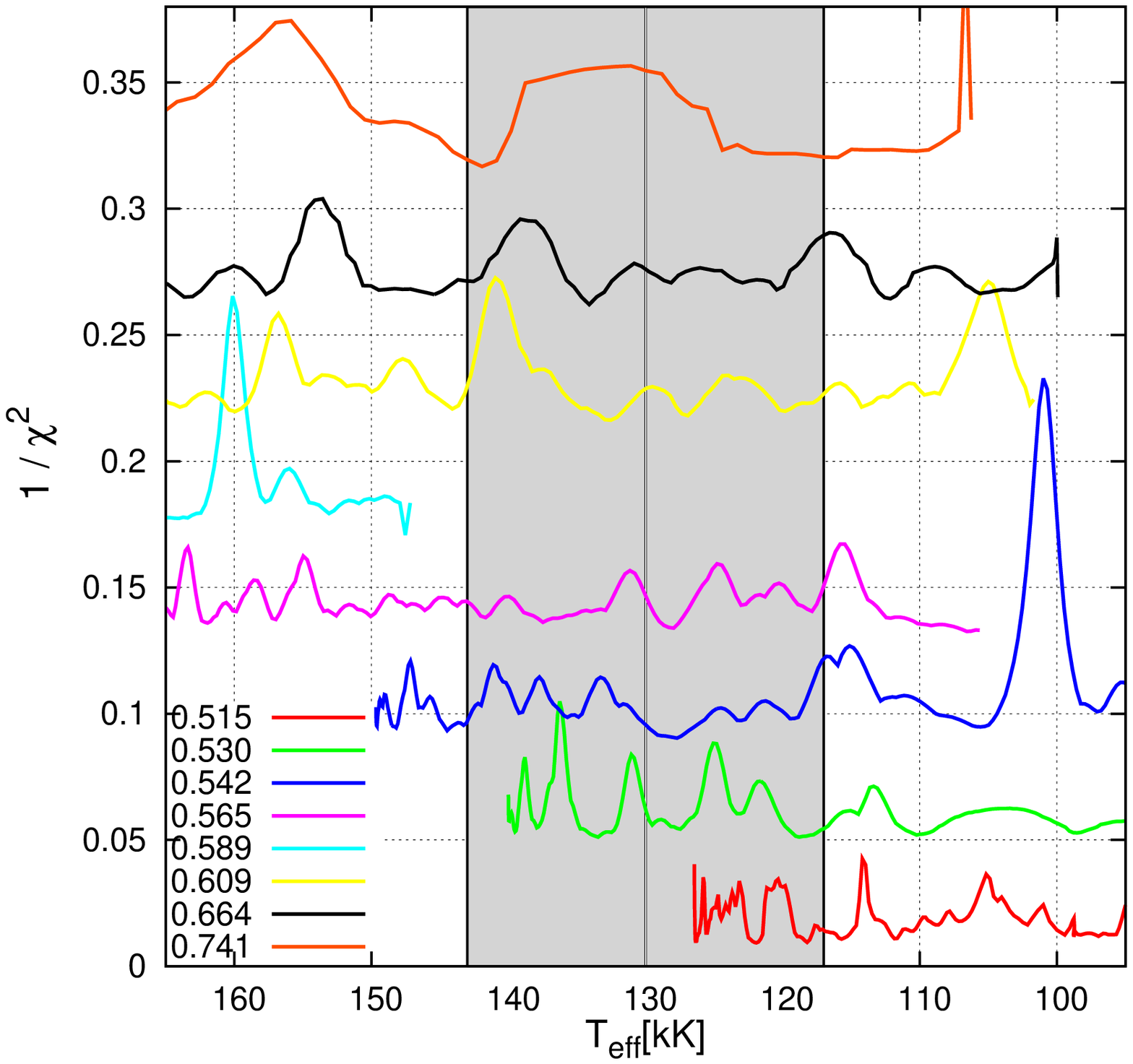}} \end{center} \caption{The
    inverse of the quality function of the period fit considering
    $\ell= 1$ modes only in terms of the effective temperature for 
    \vv\ assuming that the star is ``before
    the knee'' for three different sets of  periods (see text for
    details).  The vertical wide strip in gray depicts  the spectroscopic
    $T_{\rm  eff}$ and   its uncertainties. The
    curves have been arbitrarily shifted upward (with a step of 0.02
    for the left panel, 0.025 for the middle panel and 0.04 for the
    right panel) except for the lowest curves. It is possible to adopt a model with
    $M_{\star}= 0.530\ M_{\sun}$ [Color figure only available in the
    electronic version of the article].}  
\label{fig:ajustevvantes}
\end{figure*}

For \vv, we proceeded analogously to \J0, but discriminating the 
cases ``before the knee'' and ``after the knee''. 
We carry out the procedure considering  three
different sets of observed periods. The results of the search for the
best-fit model for this star ``before the knee'', considering only 
$\ell= 1$ theoretical modes, can be appreciated in 
Fig.~\ref{fig:ajustevvantes}. The
Fig.~\ref{fig:antes-1}, corresponds to the case of the complete set 
of periods and it shows no unambiguous asteroseismological model 
in the range of $T_{\rm eff}$ allowed by the spectroscopy. 
For the reduced list of 11 periods (see
Section~\ref{testvv}), Fig.~\ref{fig:antes-3} shows the presence of
many local maxima as well, for different masses at several values of
the effective temperature, with roughly the same amplitude of
$(\chi^2)^{-1}$. However, it is possible to find some models that may
constitute good period fits considering the range of the effective 
temperature, and, at
the same time, discarding the models with masses too high (compared to
the other mass determinations for this star). These models correspond 
to a mass of
$M_{\star}= 0.530\ M_{\sun}$ at $T_{\rm  eff} \sim 124\,500$ K and
$T_{\rm  eff} \sim 131\,000$ K. In the third case (Fig.~\ref{fig:antes-5})
the set corresponds to the 8 longest periods from the complete list (from
$1181$ s to $5682$ s), and once again, there is no unambiguous solution
in the range allowed by the effective temperature. However, it is also
possible in this case to discard models, so we may choose 
tentatively a solution for
$M_{\star}= 0.530\ M_{\sun}$ and $T_{\rm eff}$ close to the effective 
temperature of \vv. It is worth mentioning the presence of a
strong peak, present in the three cases considered, corresponding 
to the model with $M_{\star}= 0.542\
M_{\sun}$. However, in that case, the $T_{\rm  eff}$ is too low as to be
considered a solution, taking into account the constraint given by
the spectroscopy. Reversing the argument, the period fit is so good 
that this might be an indication of a very inaccurate determination 
of the effective temperature from spectroscopy for this star.
 
\begin{figure*}[ht]
  \begin{center}
    \subfigure[16 periods]{\label{fig:desp-1}\includegraphics[width=0.29\textwidth]{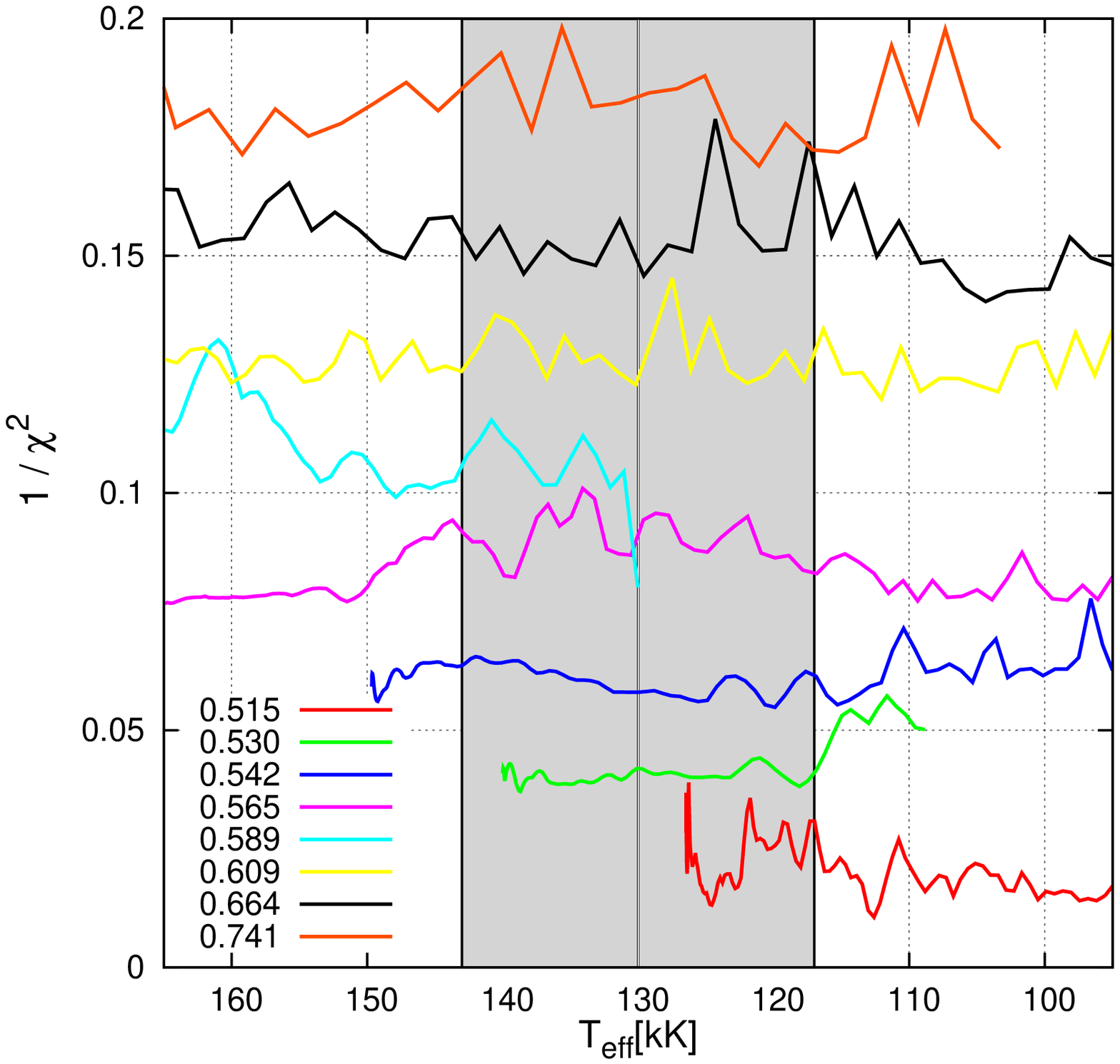}}
    \subfigure[11 periods]{\label{fig:desp-3}\includegraphics[width=0.29\textwidth]{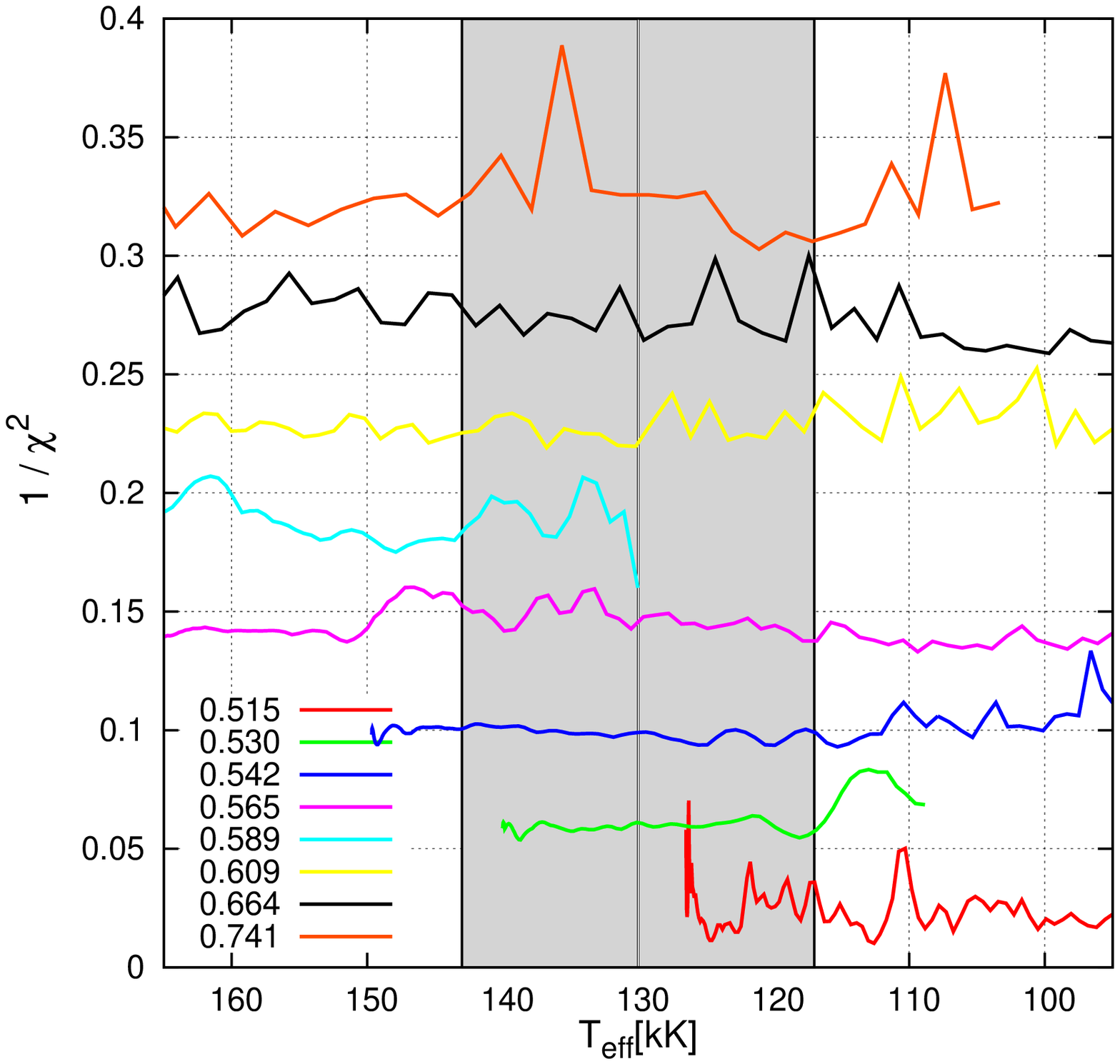}} 
    \subfigure[8 periods]{\label{fig:desp-5}\includegraphics[width=0.29\textwidth]{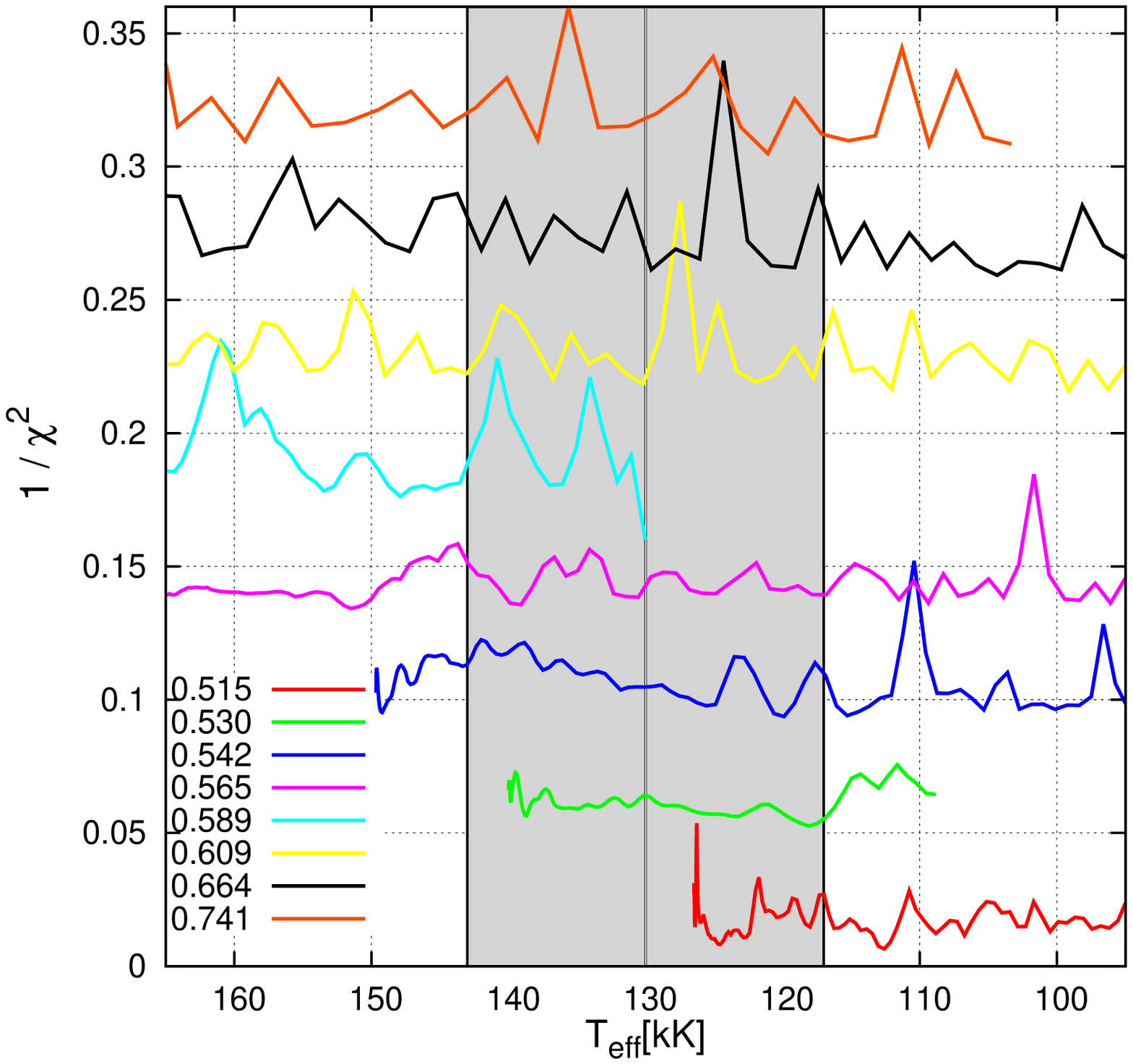}}
  \end{center}
  \caption{Same as Fig.~ \ref{fig:ajustevvantes}, but for the 
case in which \vv\ is evolving ``after the knee''.  
The curves have been arbitrarily shifted 
upward (with a step of 0.02 for the left panel and 0.04 for the middle 
and right panel) except for the lower curve.}  
\label{fig:ajustevvdesp}
\end{figure*}

 When we analyze the case ``after the knee'', we find the results shown
in Fig.~\ref{fig:ajustevvdesp}. Again, there are multiple local
maxima in the three cases. If we consider only the solutions 
that lie inside the range given for the uncertainty of the effective 
temperature, and discard those possible solutions 
associated with masses too high (as compared with the other
mass determinations for this star), we may adopt a  possible 
solution for the mass $M_{\star}= 0.515\ M_{\sun}$ at 
$T_{\rm  eff} \sim 126\,300$ K but this solution does not 
touch the value of the effective temperature of the star, so it is 
not completely reliable.

\begin{figure*}[ht!]
  \begin{center} \subfigure[16
    periods]{\label{fig:mezclaajuste-1vvantes}\includegraphics[width=0.29\textwidth]{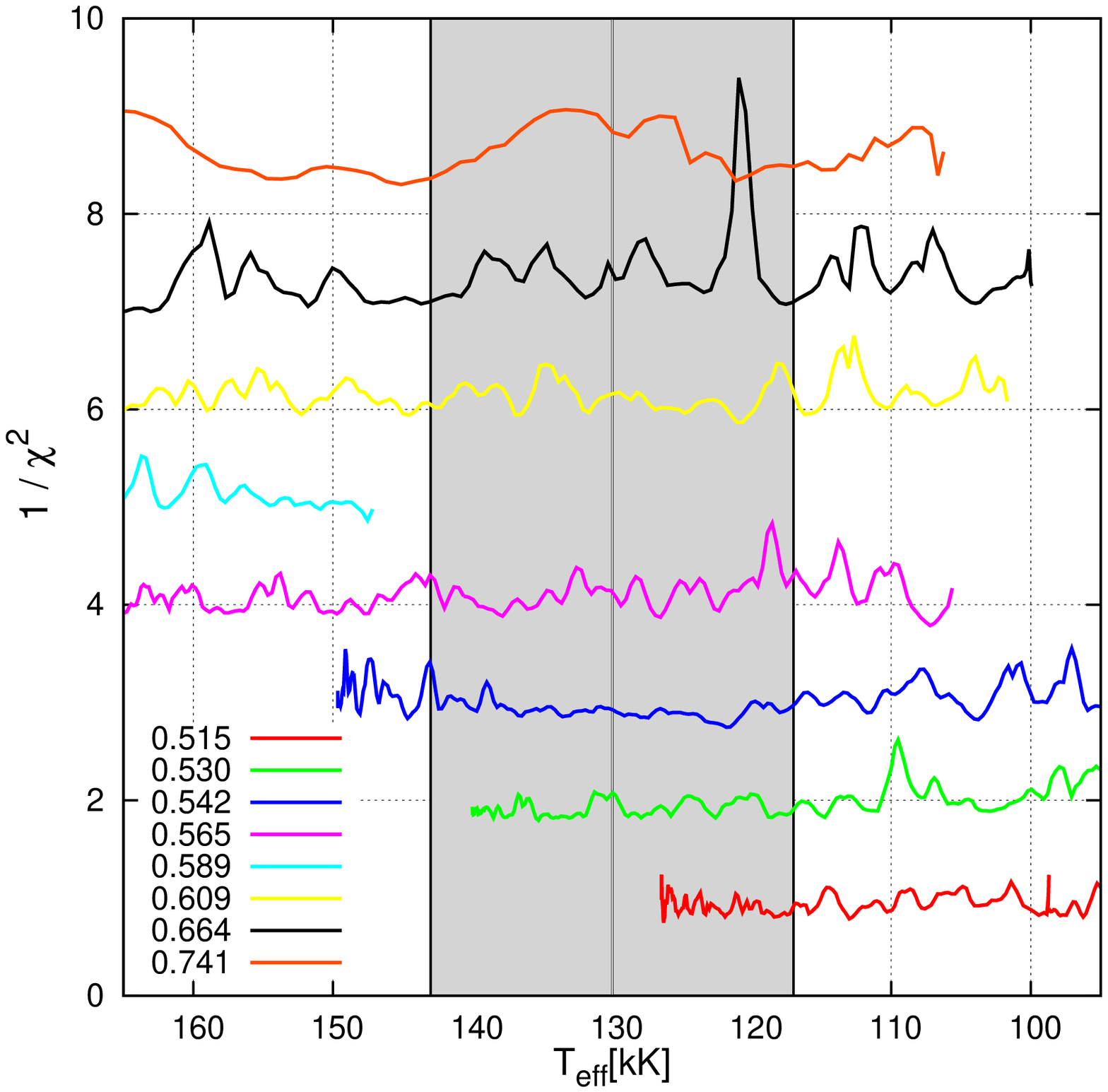}} \subfigure[11
    periods]{\label{fig:mezclaajuste-3vvantes}\includegraphics[width=0.29\textwidth]{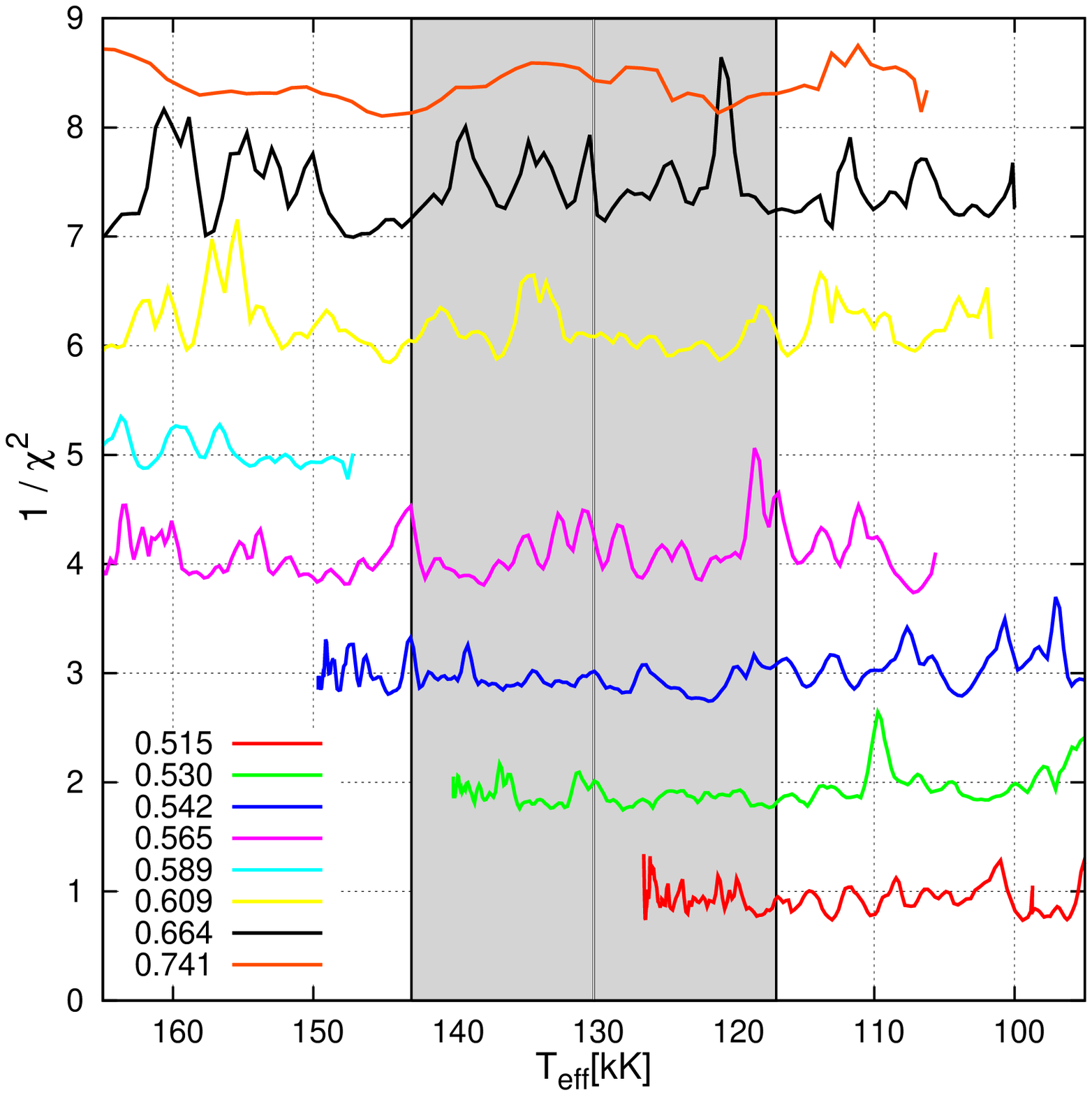}}  \subfigure[8
    periods]{\label{fig:mezclaajuste-5vvantes}\includegraphics[width=0.29\textwidth]{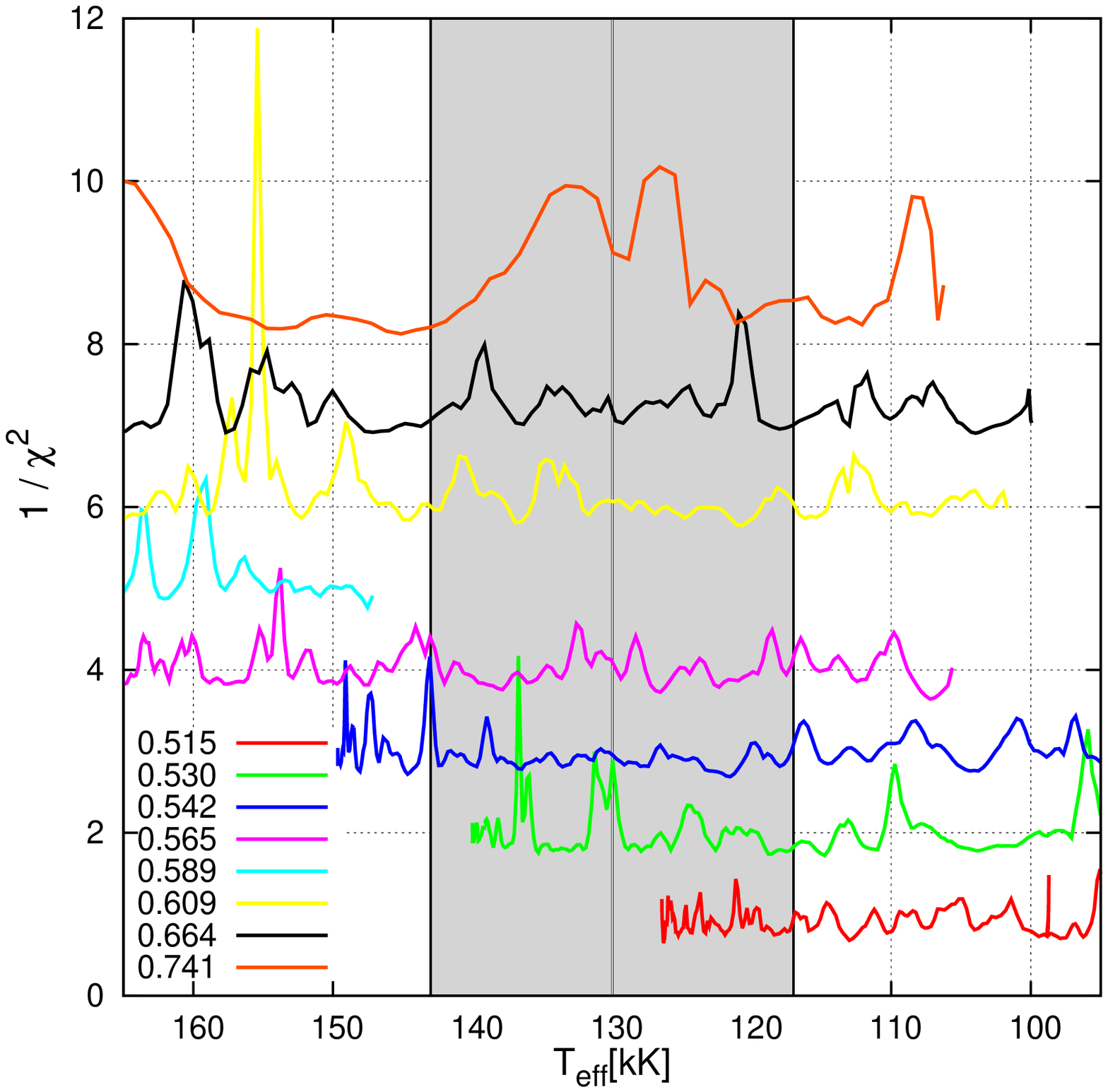}} \end{center} \caption{The
    inverse of the quality function of the period fit considering a
    mix of $\ell= 1, 2$ modes in terms of the effective temperature
    for \vv\ ``before the knee'' for three different set of  periods
    (see text for details).  The vertical gray strip depicts  the
    spectroscopic $T_{\rm  eff}$ and   its uncertainties. The curves
    have been arbitrarily  shifted upward (with a step of 1) except
    for the lowest curve. It is possible to adopt a model with    $M_{\star}= 0.530\
    M_{\sun}$  [Color figure only available in the electronic version
    of the article].}  
\label{fig:mezclaajustevvantes}
\end{figure*}

Next, we also consider the possibility that 
the periods exhibited by \vv\ are a mix of $\ell= 1$ and $\ell= 2$ modes. 
In Fig.~\ref{fig:mezclaajustevvantes} we show the results 
corresponding to the three set of periods considered for
the case in which \vv\ is ``before the knee''. For the complete set 
of periods (Fig.~\ref{fig:mezclaajuste-1vvantes}) there are several 
possible equivalent solutions, within the constraint given by the
spectroscopy, and consistent with the other determinations of the
stellar mass. In the second case (Fig.~\ref{fig:mezclaajuste-3vvantes}) 
we can appreciate a possible solution with $M_{\star}= 0.530\
M_{\sun}$ which is close to the $T_{\rm
eff}$ of \vv. In the third case 
(Fig.~\ref{fig:mezclaajuste-5vvantes}),
this possible solution is even more evident, so we 
conclude that a model with $M_{\star}= 0.530\ M_{\sun}$ and 
$T_{\rm eff} \sim 130\,000$ K may be considered as a seismological 
solution, although it is not unique. 

\begin{figure*}[ht]
  \begin{center}
    \subfigure[16 periods]{\label{fig:mezclaajuste-1vvdesp}\includegraphics[width=0.29\textwidth]{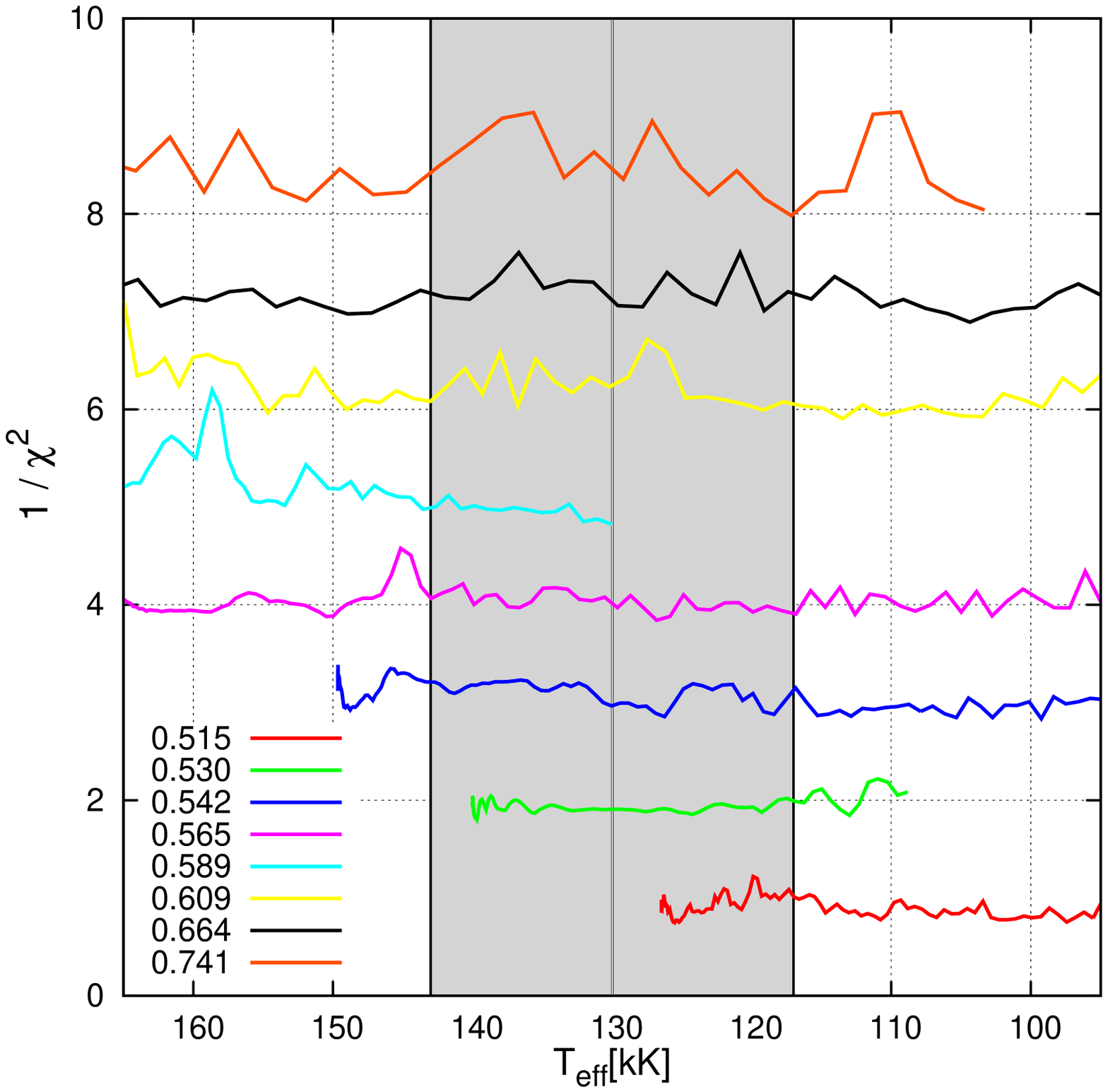}}
    \subfigure[11 periods]{\label{fig:mezclaajuste-3vvdesp}\includegraphics[width=0.29\textwidth]{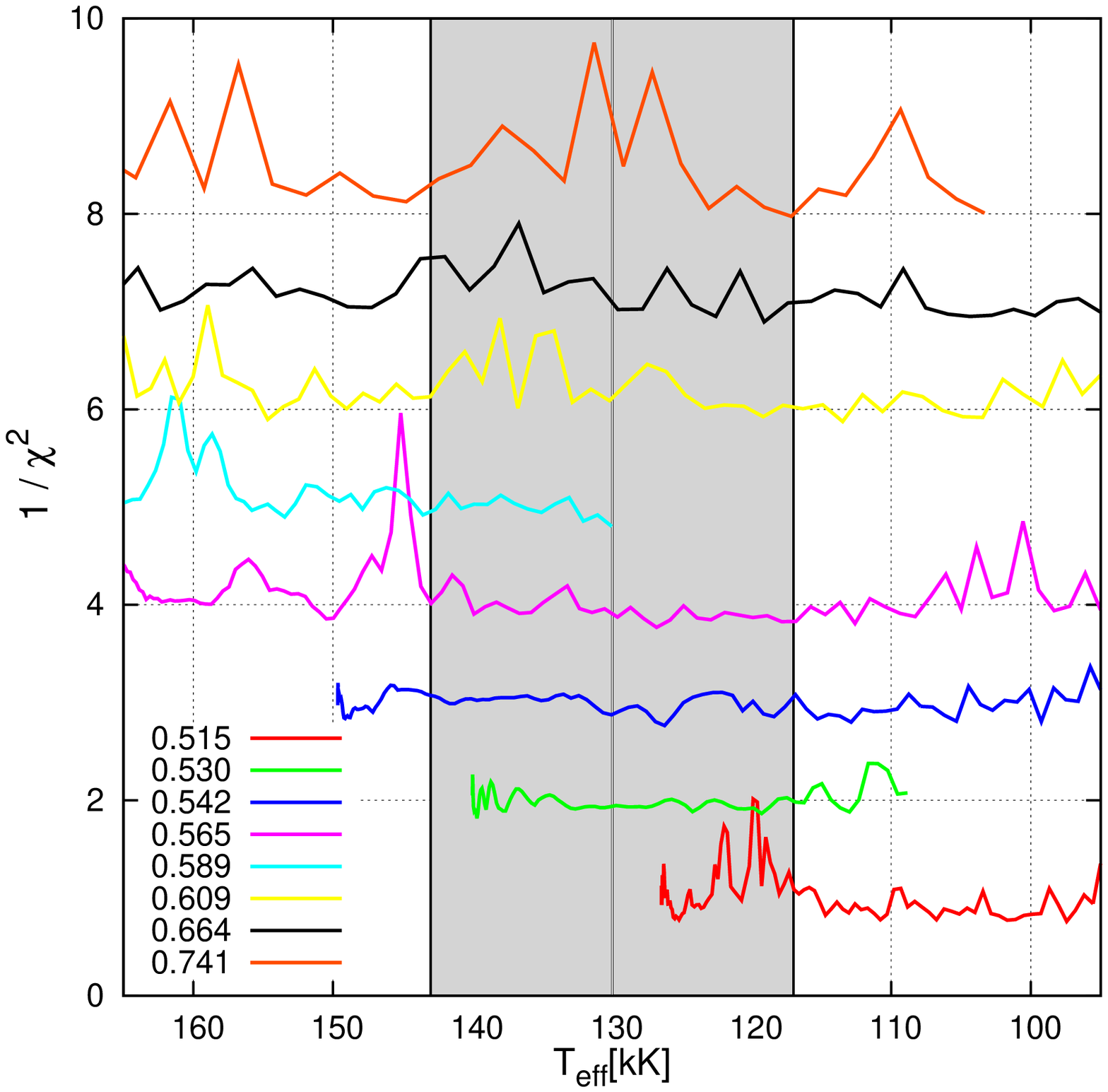}} 
    \subfigure[8 periods]{\label{fig:mezclaajuste-5vvdesp}\includegraphics[width=0.29\textwidth]{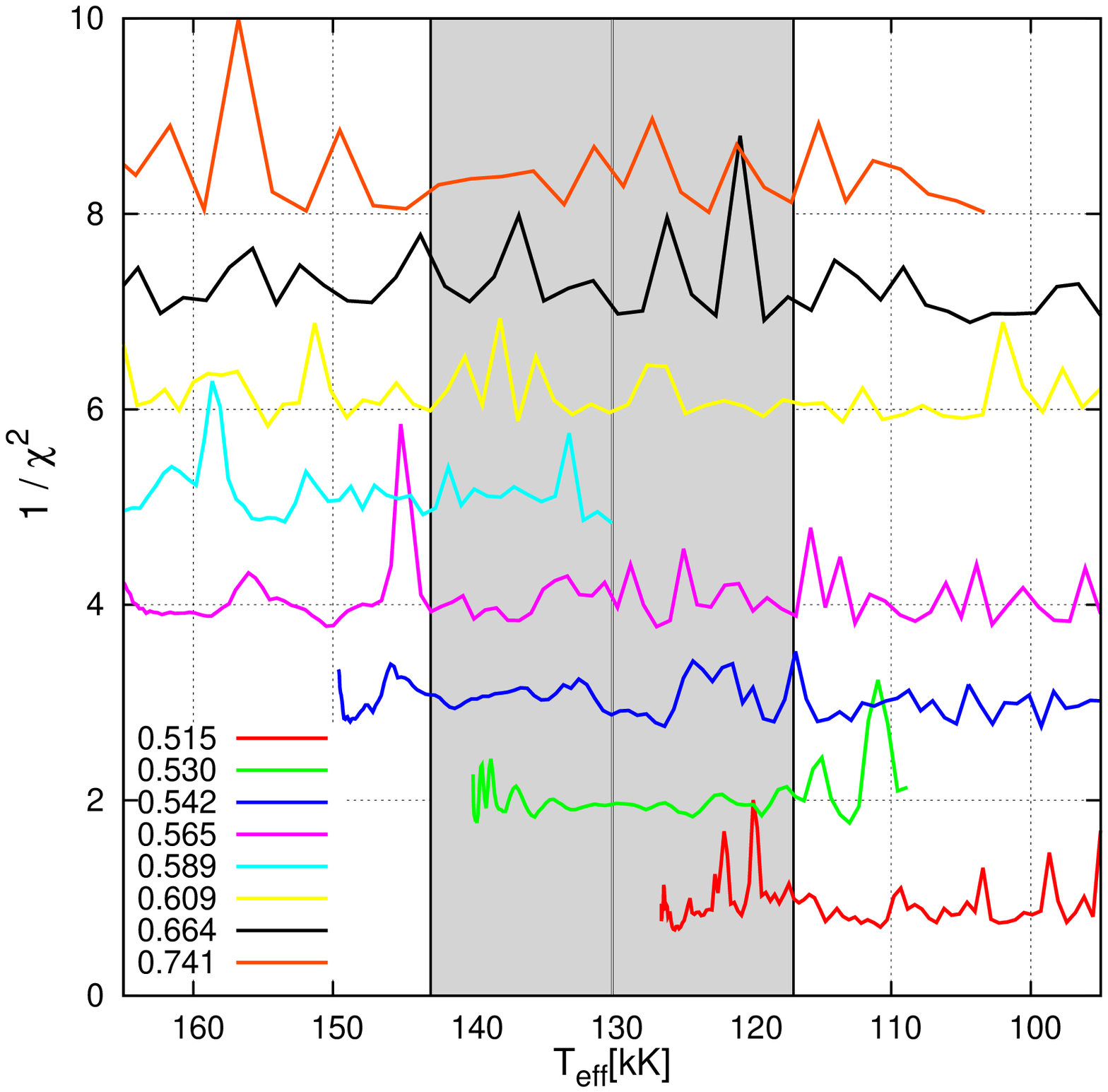}}
  \end{center}
  \caption{Same as Fig.~\ref{fig:mezclaajustevvantes}, but for 
   the case in which \vv\ is ``after the knee''.}  
\label{fig:mezclaajustevvdesp}
\end{figure*}

Finally, we considered the case in which \vv\ is ``after the knee'',
and the results are shown in  Fig.~\ref{fig:mezclaajustevvdesp}. In
the first case,  corresponding to the complete set of observed periods
(Fig.~\ref{fig:mezclaajuste-1vvdesp}),  the curves have a rather
smooth behavior, and there is not an evident  solution. The second
case, on the other hand, is  displayed in
Fig.~\ref{fig:mezclaajuste-3vvdesp}. Here,  there may be possible
solutions within the uncertainty range for $T_{\rm eff}$ that can be
adopted as representative models  of \vv\ for $M_{\star}= 0.515\
M_{\sun}$ at $T_{\rm  eff} \sim 120\,000$ K.  In the third case, which
is shown in Fig.~\ref{fig:mezclaajuste-5vvdesp}, the situation is
quite similar to the second case, and we may adopt the same solution.
As in the case for $\ell= 1$, this sequence does not 
reach the value of the effective temperature given for this star, 
so this solution may not be proper.

We close this Section by emphasizing that, all in all,  
for \vv\ we have not been able to find a \emph{clear
and unambiguous} seismological solution on the basis of 
our set of PG 1159 evolutionary models. This prevent us to adopt 
a representative asteroseismological model for this star (as we did 
for \J0), and thus to  infer its internal structure.

\section{Rotation of \J0}  
\label{splitting} 

As noted by \cite{2012MNRAS.426.2137W}, there is a probable triplet of
frequencies in the 2007 data set at $2372.6$ $\mu$Hz, $2387.2$ $\mu$Hz, 
and $2401.4\mu$Hz ($421.48$ s, $418.90$ s,  and $416.42$ s,
respectively). This makes feasible a first order analysis of 
the  rotational splitting in order to derive an estimate  
the rotation period, $P_{\rm rot}= 1/\Omega$, of \J0.

In absence of rotation, each eigenfrequency of a nonradially pulsating star 
is ($2 \ell + 1$)-fold degenerate. Under the assumption of slow rotation, 
which is the case for most of white dwarfs and pre-white dwarfs 
 \citep{2008PASP..120.1043F,2009Natur.461..501C,2010A&ARv..18..471A}, 
the perturbation theory can be applied 
to the first order. It turns 
out that the degeneracy of frequencies is lifted, and each component 
of the resulting multiplet can be calculated as: 
$\sigma_{k,\ell, m}(\Omega)= \sigma_{k,\ell}(\Omega= 0)
+ \delta \sigma_{k,\ell, m}$ \citep{1949ApJ...109..149C}. 
If rotation is rigid ($\Omega$= constant)
the first-order corrections to the frequency can be expressed 
as: $\delta \sigma_{k,\ell, m}=
-m \Omega (1-C_{k,\ell})$, where $m= 0, \pm 1, \cdots , \pm \ell$, and 
$C_{k\ell}$ are  coefficients that depend on the details of the stellar
structure and the eigenfunctions, and can be obtained in the non-rotating
case. When an asteroseismological model is found for the star under
study, such coefficients are available (i.e. they result from the
pulsation calculations) and so, it is our case.  For \J0, the 
asteroseismological model is characterized by $T_{\rm eff}= 91\,255$ K and 
$M_{\star}= 0.542 M_{\sun}$ (Table \ref{tab:modeloj0}).
If we associate the periods of \J0\ at $416.42$ s and $421.48$ s to 
the components of a rotational triplet with $m= +1$ and $m= -1$ 
respectively, and assume the period $418.90$ s as the central 
component ($m= 0$) (see Table \ref{table1}), 
then $\Delta \sigma= \sigma(m= +1) - \sigma(m= -1)= 28.8\ \mu$Hz is the
frequency spacing between the extreme components of the triplet. Thus,
$\delta \sigma= \Delta\sigma / 2= 14.4\ \mu$Hz. The corresponding value
for the coefficient $C_{k,\ell}$ is $0.4936$, corresponding to the
theoretical period closest to the observed period with $m= 0$ of the 
asteroseismological model for \J0. Then,
$\Omega= 28.4360\ \mu$Hz, leading to a rotation period of $P_{\rm
rot}= 1/\Omega= 0.407\ $d.

\section{Summary and conclusions}  
\label{conclusions}  
  
In  this  work,  we  presented  a detailed asteroseismological   study
of  the pulsating PG 1159 stars \J0\ and \vv, aimed at determining
the internal structure and evolutionary stage of these pulsating
stars.  Our analysis is  based on the  fully evolutionary PG 1159
models of \cite{2005A&A...435..631A},  \cite{2006A&A...454..845M}  and
\cite{2006A&A...458..259C}. The observational
data employed for this study was based on the observed periods
reported in the works of \cite{2012MNRAS.426.2137W} for \J0,
and \cite{2006A&A...454..527G} for \vv. Employing the spectroscopic
data from \cite{2006A&A...454..617H} for \J0 and
\cite{2006PASP..118..183W} for \vv, we inferred a value for the spectroscopic
mass of both stars. The results are $M_{\star}= 0.543 M_{\sun}$ 
for \J0 and $M_{\star}= 0.529\ M_{\sun}$ for \vv\ (see
Fig.~\ref{grave-obs}).

\begin{table*}[t]
\begin{center}
\caption{Stellar  masses  for all of the intensively studied pulsating  
         PG 1159 stars. All masses are in solar units.}
\begin{tabular}{lccclc}
\hline 
\hline
\noalign{\smallskip}
Star & $\Delta \Pi_{\ell}^{\rm a}$ &$\overline{\Delta \Pi_{\ell}}$ &  Period fit & Pulsations    & Spectroscopy\\
     &                             &                             &            & (other works) & \\
\noalign{\smallskip}
\hline
\ngc       & 0.571$^{\rm a}$        & 0.576$^{\rm a}$ &  ---             & 0.55$^{\rm j}$ (asymptotic analysis) & 0.56 \\ %NGC1501
\rxj       & 0.568$^{\rm b}$        & 0.560$^{\rm b}$ &  0.565$^{\rm b}$ & 0.56$^{\rm h}$ (asymptotic analysis) & 0.72 \\ %RXJ2117
\pp        & 0.577--0.585$^{\rm d}$ & 0.561$^{\rm d}$ &  0.565$^{\rm d}$ & 0.59$^{\rm i}$ (asymptotic analysis) & 0.54 \\ %PG 1159
\pr        & 0.627$^{\rm a}$        & 0.578$^{\rm a}$ &  0.589$^{\rm a}$ & 0.61$^{\rm e}$ (period fit)          & 0.55 \\ %PG2131
\pt        & 0.597$^{\rm a}$        & 0.566$^{\rm a}$ &  0.542$^{\rm a}$ & 0.57$^{\rm g}$ (asymptotic analysis) & 0.53 \\ %PG1707
\pg        & 0.625$^{\rm c}$        & 0.567$^{\rm c}$ &  0.566$^{\rm c}$ & 0.69$^{\rm f}$ (asymptotic analysis) & 0.53 \\ %PG0122
\rrr       & ---                   & ---            &  0.556$^{\rm k}$ &  ---                               & 0.52  \\ %SDSS J0754
\J0        & 0.569$^{\rm l}$        & 0.535$^{\rm l}$ &  0.542$^{\rm l}$ &  ---                               & 0.54 \\ %SDSSJ0349  
\vv        & 0.523$^{\rm l}$        & 0.528$^{\rm l}$ &  0.523$^{\rm l}$ &  ---                               & 0.51-0.61 \\ %VV47  
\hline
\hline
\end{tabular} 
\label{tabla-masas}
\end{center}

{\footnotesize  Notes: 
$^{\rm a}$\citet{2009A&A...499..257C}.   
$^{\rm b}$\citet{2007A&A...461.1095C}.  
$^{\rm c}$\citet{2007A&A...475..619C}.  
$^{\rm d}$\citet{2008A&A...478..869C}.  
$^{\rm e}$\citet{2000ApJ...545..429R}.  
$^{\rm f}$\citet{2007A&A...467..237F}.
$^{\rm g}$\citet{2004A&A...428..969K}.   
$^{\rm h}$\citet{2002A&A...381..122V}.
$^{\rm i}$\citet{2008A&A...477..627C}. 
$^{\rm j}$\citet{1996AJ....112.2699B}.
$^{\rm k}$\citet{2014MNRAS.442.2278K}.
$^{\rm l}$This work.}
\end{table*}

Next, we determined the observed period spacing for both stars, 
employing three different and independent
tests.  We found $\Delta\Pi^{\rm O}_{\ell= 1}= 23.4904 \pm 0.07741$ s
for \J0 and $\Delta\Pi^{\rm O}_{\ell= 1}= 24.2015 \pm 0.03448$ s
for \vv\ (see Sect. \ref{estimation}).  Then, making use of the strong
dependence of the period spacing of pulsating PG 1159 stars on the
stellar mass, we derived the mass for both stars under
study. First, it was achieved by comparing the observed period
spacing with  the asymptotic  period spacing of our models (which is
an inexpensive approach, since it that does  not involve pulsation
computations). We obtained $M_{\star}= 0.569^{+0.004}_{-0.002}\
M_{\sun}$ for \J0. For \vv, we derived 
$M_{\star}= 0.526^{+0.007}_{-0.005}\ M_{\sun}$ if the star  
is ``before the knee'', and $M_{\star}=
0.520^{+0.002}_{-0.005}\ M_{\sun}$ for \vv, if instead it is ``after the knee''
(see Sect. \ref{sect-aps}). A second estimate of $M_{\star}$, 
based on the comparison of the  observed period spacings with 
the average   of the computed period spacings 
(an approach that requires of detailed
period  computations), gives  $M_{\star}= 0.535 \pm 0.004\
M_{\sun}$ for \J0, and $M_{\star}= 0.532^{+0.004}_{-0.007}\ M_{\sun}$
(``before the knee'') and $M_{\star}= 0.524^{+0.002}_{-0.001}\ M_{\sun}$ 
(``after the knee'') for \vv\
(see Sect. \ref{sect-psp}). A third determination was achieved by
carrying out period-to-period fits, consisting in searching for 
models  that best reproduce the  individual observed periods of each star. 
The period fits were made on a grid of PG 1159 models with a fine resolution in
stellar mass and a much finer grid in effective temperature and
considering  $g$ modes with $\ell= 1$ and $\ell= 2$.  In the case
of \J0, we were able to find an asteroseismological model with
$M_{\star}= 0.542\ M_{\sun}$ and $T_{\rm eff}= 91\, 255\ $K (for
$\ell= 1$ $g$ modes) based on the constraint given by the 
spectroscopy (see \ref{searchingj0}). The search for a period fit 
for modes $\ell= 1$ and $\ell= 2$ simultaneously did not result 
in an asteroseismological solution, thus indicating 
that the periods exhibited by  this star are associated only to 
$\ell= 1$ modes.  In the case of \vv\   
there is not a clear and unambiguous solution (see Sect. \ref{searchingvv}), 
which unfortunately prevents us to present a representative 
aseroseismological model for this star and to extract seismic 
information of its internal structure.
Finally, for \J0, once we
adopt the model with $M_{\star}= 0.542\ M_{\sun}$ as the 
asteroseismological model for this star, we were able to determine the
rotation period employing the observed triplet of frequencies
associated with the period $418.90\ $s ($m= 0$) (see
Sect. \ref{splitting}). We found a rotation period of $P_{\rm rot}=
1/\Omega= 0.407$ d.

In Table \ref{tabla-masas} we show a compilation of the mass determinations
carried out for the most studied pulsating PG 1159 stars, analyzed 
on the basis of our set of fully evolutionary models. For the case of 
\vv, the values quoted for the mass result from averaging the 
two estimates obtained in this paper considering that the star is evolving 
 before and after the evolutionary knee.

In summary, we were able to find an excellent agreement between our
estimates for the stellar mass of both \J0\ and \vv, something that shows
the great internal consistency  of our analysis. The fact that we were
able to find an asteroseismological model for \J0 implies that we have
additional information about this star, such as the stellar radius,
luminosity and gravity  (see Table \ref{tab:modeloj0}). On the other
hand, as expected, the determination of the rotation period shows a good agreement
with the one estimated by \cite{2012MNRAS.426.2137W} of $P_{\rm rot}=
1/\Omega= 0.40 \pm 0.01 $d, and it is also in line with the values
determined for other white dwarf and pre-white dwarf stars
(see \cite{2008PASP..120.1043F}, Table 4). This result reinforces the
belief that pre-white dwarf stars are slow rotators. We mention that
it would be valuable to repeat these calculations using an independent set of
evolutionary tracks of PG 1159 stars to add robustness to our results. This is
beyond the scope of this paper.

The present paper constitutes a further step in a series of  studies
done by our group aiming at studying the internal structure and
evolution status of  pulsating PG 1159-type stars through the tools of
asteroseismology.  From the results presented in this paper for
the pulsating PG 1159 stars \J0\ and \vv, it is evident once again 
the power of this approach, in
particular for  determining the stellar mass with an unprecedented
precision.

%------------------------------------------------------------------------  
  
\begin{acknowledgements}
We wish to thank our anonymous referee for the constructive
comments and suggestions that greatly improved the original version of
the paper.
Part of this work was supported by AGENCIA through the 
Programa de Modernizaci\'on Tecnol\'ogica BID 1728/OC-AR, and by the 
PIP 112-200801-00940 grant from CONICET. This research made use 
of NASA Astrophysics Data System.
\end{acknowledgements}

\bibliographystyle{aa}
\bibliography{biblio}

\end{document}